# Probing the electronic structure and photophysics of thiophene–diketopyrrolopyrrole derivatives in solution

Daniel W. Polak,[1] Mariana T. do Casal,[2] Josene M. Toldo,[2] Xiantao Hu,[3] Giordano Amoruso,[1] Olivia Pomeranc,[1] Martin Heeney,[3] Mario Barbatti,[2,4] Michael N. R. Ashfold,[1] and Thomas A. A. Oliver[1]*

[1] School of Chemistry, Cantock's Close, University of Bristol, Bristol, BS8 1TS, UK

[2] Aix Marseille Université, CNRS, ICR, Marseille, France

[3] Department of Chemistry and Centre for Processable Electronics, Imperial College London, White City Campus, London, W12 0BZ, UK

[4] Institut Universitaire de France, 75231, Paris, France

*Author for correspondence: tom.oliver@bristol.ac.uk

## Abstract

Diketopyrrolopyrroles are a popular class of electron-withdrawing unit in optoelectronic materials. When combined with electron donating side-chain functional groups such as thiophenes, they form a very broad class of donor-acceptor molecules: thiophene-diketopyrrolopyrroles (TDPPs). Despite their widescale use in biosensors and photovoltaic materials, studies have yet to establish the important link between the electronic structure of the specific TDPP and the critical optical properties. To bridge this gap, ultrafast transient absorption with 22 fs time resolution has been used to explore the photophysics of three prototypical TDPP molecules: a monomer, dimer and polymer in solution. These studies show that the photophysics of these molecular prototypes under visible photoexcitation are determined by just two excited electronic states, having very different electronic characters (one is optically bright, the other dark), their relative energetic ordering and the timescales for internal conversion from one to the other and/or to the ground state. The underlying difference in electronic structure alters the branching between these excited states and their associated dynamics. In turn, these factors dictate the fluorescence quantum yields, which are shown to vary by ~1–2 orders of magnitude across the TDPP prototypes investigated here. The fast non-radiative transfer of molecules from the bright to dark states is mediated by conical intersections. Remarkably, wavepacket signals in the measured transient absorption data carry signatures of the nuclear motions that enable mixing of the electronic-nuclear wavefunction and facilitate non-adiabatic coupling between the bright and dark states.

# 1. Introduction

Molecules with alternating electron rich and electron deficient moieties are known as donor-acceptor (DA) chromophores. DA molecules are an important category of material due to their tuneable and low energy band gaps which originate from alternating electron densities across the molecular framework.[1,2] Diketopyrrolopyrroles (DPPs) are a popular class of electron acceptor unit, that display a range of excited state dynamics.[3–6] The judicious choice of connecting donor structures between DPP repeat units can have a major influence on the associated fluorescence quantum yield, non-radiative rates, probability of charge-transfer, and excited state lifetimes of the resulting molecule. This semi-empirical tunability of molecular parameters means DPP-containing systems can support processes such as exciton dissociation, long-range charge-separation, singlet fission[7–16] and even symmetry-breaking charge-transfer, which has led to their application in a range of optical applications such as biosensors[17–19] or photoelectronic/photovoltaic devices.[3,20–29]

Due to their wide-ranging applications, many studies have investigated the structure–function relationship of DPP-containing thin films.[30–34] A greater understanding of how the molecular structure and deposition method affects film morphology has allowed for the creation of 12% efficient solar cells with DPP-based materials as the active photovoltaic component.[3,24–26,35–37] The interdependence of microscopic and mesoscopic phenomena in such solid samples hampers the development of reliable mechanistic insights into their operation. Solution-phase studies of DPP-containing systems offer a way of circumventing morphological complexities associated with thin films and allow for systematic investigations of the excited state dynamics under conditions more amenable to comparison with electronic structure calculations. In the solution phase studies conducted to date, changes in chemical structure and environment have reportedly led to drastic changes in photodynamical behaviour.[9,12,13,15,38–41]

DPP-containing molecules typically have nanosecond excited state lifetimes with appreciable fluorescence quantum yields (>50%).[7,8,12,14,19,40] The corresponding thiophene-diketopyrrolopyrrole (TDPP) family of molecules are no exception to this observation.[12] Counter to this general trend, a dimer of TDPP bridged by a vinylene linker (TDPP-v-TDPP) has no measurable fluorescence and a very short excited state lifetime.[38] Many experimental and computational studies of other DPP-containing monomers,[7,8,12,14,40] polymers,[23,31,39] and dimers[9,13,41] have failed to identify similar behaviour to that of the vinylene linker containing dimer. This inconsistency in the excited state dynamics motivated the present fundamental, ‘bottom up’ investigation of the electronic structure and photophysics of the prototypical TDPP-based systems. The chemical structures of the compounds studied experimentally, along with their respective absorption spectra in toluene solution, are given in Fig. 1.

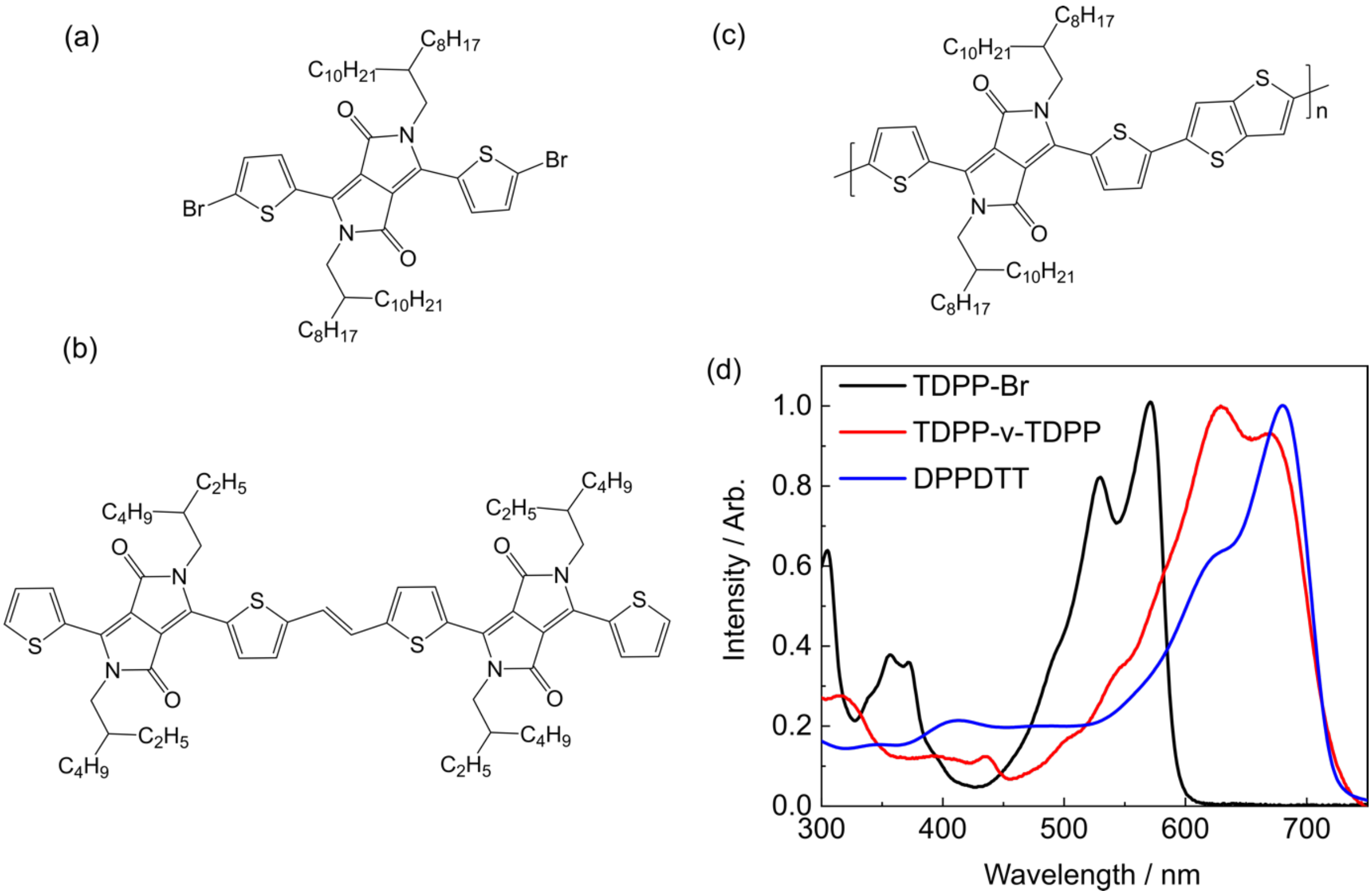

**Fig. 1** Chemical structures of (a) monomer TDPP-Br, (b) dimer TDPP-v-TDPP and (c) polymer DPPDTT. (d) Steady state absorption spectra of the three studied compounds in toluene solution.

The present study used transient absorption (TA) spectroscopy to unravel the strikingly different dynamics of a monomer (TDPP-Br), dimer (TDPP-v-TDPP) and polymer (DPPDTT[42]). Experimental results are compared to recent multireference configuration interaction density functional theory (DFT/MRCI) and spin-flip time-dependent DFT (SF-TD-DFT) calculations of TDPP and TDPP-v-TDPP molecules.[43] These studies demonstrate that two photoexcited states, the optically ‘bright’ and ‘dark’ excited states, are key to determining the visible photochemistry of the different TDPP systems. The observed photophysical differences are shown to be dominantly governed by the energetic re-ordering of the first two excited states, which greatly impacts the excited state dynamics, changing them from predominantly fluorescent decay (monomer) to non-fluorescent (dimer). Further, transient absorption data contain signatures of the nuclear motions responsible for driving ultrafast conversion between the critical bright and dark electronic states- a phenomenon rarely observed in ultrafast condensed phase dynamics of polyatomic molecules.

## 2. Results and Discussion

## 2.1 TDPP-Br

### 2.1.1. Photophysics

The steady state absorption and fluorescence spectra of TDPP-Br are displayed in Fig. 2(a). TA spectroscopy using broadband 22 fs pump pulses (centred at 580 nm, FWHM ~ 80 nm) and visible/near-IR continuum probe pulses was used to follow the excited state dynamics of TDPP-Br in toluene solution. The spectrally resolved TA data displayed in Fig. 2(b) are plotted in terms of differential transmittance ($\Delta T/T$) for four representative pump-probe time delays, exhibiting negative features centred at probe wavelengths ~450, 710 and 750 nm and a prominent set of positive peaks between 500 and 680 nm. The profile of the positive features has a similar spectral contour to a weighted sum of the parent absorption and fluorescence spectra. Given this close resemblance, it is logical to associate the positive transient features at $\lambda < 580$ nm with depletion of the ground state population by the pump pulse corresponding to a ground state bleach (GSB) and, at $\lambda > 580$ nm, to stimulated emission (SE) from a photoexcited state. The negative features at wavelengths longer than 650 nm are assigned to excited state absorption (ESA) transients.

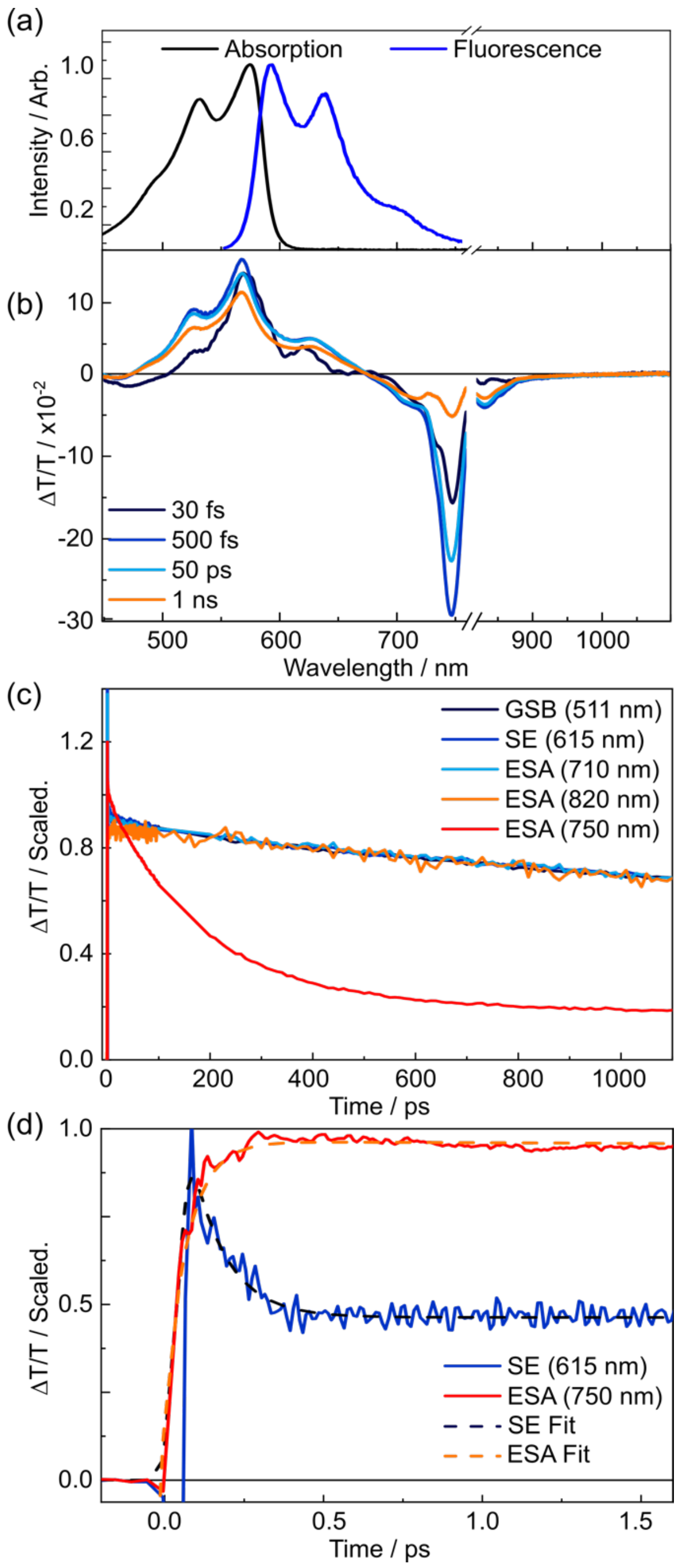

**Fig. 2** Steady state and time resolved spectroscopic characterisation of TDDP-Br in toluene solution. (a) Linear absorption and fluorescence spectra. (b) Transient absorption spectral slices measured at four illustrative pump-probe time delays. Probe wavelengths close to the 800 nm fundamental have been removed. (c) Normalised late time dynamics associated with five different probe wavelengths. (d) Normalised early time dynamics (out to 1.6 ps) and overlaid fits to data (dashed lines) associated with 615 nm and 750 nm probe wavelengths.

The negative transient features were assigned by examining the kinetics associated with the maximal intensity of each signal, which are displayed in Fig. 2(c) and scaled for comparison.

Two distinct decay regimes were found within the 1.1 ns time measurement window: a fast dynamical component occurring on the timescale of a few picoseconds, and a much slower component spanning hundreds of picoseconds through to nanoseconds. The kinetic traces presented in Fig. 2(c) for the positive GSB and SE features and for the negative ESA transients at 710 and 820 nm all show very similar, if not identical, kinetics. It was not possible to accurately determine the lifetime of the long-lived component, as the 5 ns time constant returned by fitting is far longer than the optical delay window, but this value is in accord with a previous study of (Br-free) TDPP.[9]

The most prominent ESA transient, peaking at 750 nm, decays far more rapidly than all other negative features. The first 1.6 ps of dynamics for the 615 nm SE and the 750 nm ESA signals, overlaid with fits to the data, are shown focusing on the earliest time delays in Fig. 2(d). Section 3 of the Electronic Supplementary Information (ESI) gives further results from the kinetic analysis. During the earliest time delays, the 615 nm SE signal was observed to decay with a time constant $\tau = 103 \pm 30$ fs (that accounted for 59% of the total signal amplitude), and the 750 nm ESA feature was found to rise with a matching time constant ($105 \pm 5$ fs). The strong agreement between the time constants of the transient species responsible for these features implies direct conversion from the initial photo-prepared population into another excited state, the latter with a distinct ESA spectrum centred at 750 nm. After this initial decay/rise, the 615 nm SE feature decays slowly, on an identical timescale to the GSB recovery, whereas the 750 nm ESA feature decays more rapidly. The latter decay is well described by a single exponential (with $\tau = 175 \pm 5$ ps (73 %)), together with an additional long-time offset ($\tau \sim 5$ ns, 27 %). We note that the 5 ns lifetime component is similar to the long-lived population observed in Fig. 2(c), suggesting this component is attributable to the long wavelength tail of the same parent ESA revealed in the kinetic fits to ESA signals at 710 and 820 nm.

The prompt but incomplete population transfer from the photo-prepared state to another excited state implies that multiple electronic states contribute to the very early time dynamics of TDPP-Br. To gain further insights into the electronic structure of TDPP-Br, a discussion and comparison with recent DFT/MCRI calculations by Casal *et al.* of very similar molecules to those studied experimentally is helpful[43] (see Table 1). To reduce computational costs, the prior computational study truncated the pendant branched alkyl chains (see Fig. 1) with methyl groups and the Br atoms were replaced with H atoms (henceforth referred to as simplified structures); moreover, the calculations were performed on isolated molecules (thus, solvation effects are absent). In this previous study, the simplified structures of TDPP-Br and TDPP-v-TDPP were optimized with DFT in the ground state and SF-TD-DFT in the excited singlet states. Subsequent single point energy calculations of the resulting structures were carried out using DFT/MRCI. The validity of truncating the alkyl chains and the effects of this simplification on the ground state geometry/symmetry are explored with additional DFT and TDDFT calculations discussed in the ESI section 7. For the sake of simplicity, we refer to electronic states as labelled by their corresponding symmetry in the $C_{2h}$ point group throughout.

Geometry optimisation returned a planar ground ($S_0$) state equilibrium geometry for the simplified TDPP, with $A_g$ electronic symmetry (in the $C_{2h}$ point group), as depicted in Fig. S6 of the ESI. The lowest excited singlet state ($S_1$) also has a planar equilibrium geometry but $B_u$ symmetry. Single point calculations on these structures returned corresponding vertical and adiabatic excitation energies of 2.34 and 2.19 eV, respectively (equivalent to ~530 and ~566 nm) and identified the $S_1$ state as optically 'bright', formed predominantly via a $\pi^* \leftarrow \pi$ excitation. Henceforth, the ground electronic state will be described interchangeably as the $S_0$ or $1A_g$ state, and the first singlet excited state of TDPP (and TDPP-Br) will be referred to as the $1B_u$ state.

The $S_2$ state of the simplified TDPP has $A_g$ symmetry and a vertical excitation energy of 3.27 eV (~ 379 nm). The $S_2$–$S_0$ transition dipole moment is predicted to be zero; the $S_2$ state is thus optically 'dark'. This state has a large contribution from two-electron promotions (*i.e.*, it has a substantial doubly excited character, %D in Table 1). Minimising the structure of $2A_g$ state confirms that it too has a planar equilibrium geometry and a subsequent single point energy calculation yields an adiabatic excitation energy of 2.92 eV (equivalent to a wavelength ~ 425 nm).

**Table 1** Calculated vertical (VEE) and adiabatic (AEE) excitation energies, oscillator strengths (*f*) and percentage of two electron (double) excitation character (%D) for the simplified TDPP monomer. Reported excitation energies were computed at the DFT/MRCI/SV(P) level of theory and reproduced from ref. [43].

| State | VEE / eV (nm) | AEE / eV (nm) | $S_n \leftarrow S_0$ $f$ | %D ($S_0$ min) |
|---|---|---|---|---|
| $S_1$ ($1B_u$) | 2.34 (530) | 2.19 (566) | 0.60 | 7.4 |
| $S_2$ ($2A_g$) | 3.27 (379) | 2.92 (425) | 0.00 | 39.4 |

Given that theory predicts the optically 'dark' $2A_g$ state of the simplified variant of TDPP-Br to be at higher energy than the 'bright' $1B_u$ state in the vertical Franck-Condon region,[43] the most likely explanation for the early time experimental data (Fig. 2(d)) is that the short wavelength edge of the broadband pump pulse populates vibrationally excited levels of the $1B_u$ state with sufficient energy to access the $2A_g$ state. At some point subsequent to vibrational cooling, the $2A_g$ and $1B_u$ state populations are no longer able to interconvert, locking-in the respective populations. This is reinforced by the wavelength dependence of the fluorescence quantum yield ($\Phi_F$) of TDPP-Br: as the excitation wavelength is tuned from 530 to 500 and then 475 nm, the $\Phi_F$ decreases by 13% (from 61, to 57 and then 48%, respectively), consistent with the hypothesis that an alternative (non-Kasha type) decay pathway becomes more prominent at shorter wavelengths.

Thus, we arrive at a picture wherein photoexcitation of TDPP-Br populates a broad spread of $1B_u$ state vibrational levels. Some of the most highly excited vibrational states of the $1B_u$ state couple to $2A_g$ state levels ($\tau$ = 105 fs), but the remainder vibrationally cool to the $1B_u$ state minimum from whence they radiatively decay ($\tau$ ~5 ns). Population that relaxes into the $2A_g$ minimum is tracked via the 750 nm ESA feature, which is found to decay on a faster ($\tau$ = 175 ps) timescale, with no associated recovery of the ground state bleach. The decay of this ESA is mirrored by the growth of a small negative transient on the blue edge of the probe window (400–475 nm), indicating the 'dark' state decays to a further excited state rather than the ground state. Hartnett *et al.* reported intensity in a similar wavelength region in previous triplet sensitisation studies of a (Br-free) TDPP in solution and ascribed the feature to triplet excited state absorption.[12] We similarly assign the $\tau$ ~175 ps decay component in the TDPP-Br data to intersystem crossing from the $2A_g$ state into the triplet manifold. Hartnett *et al.* did not observe triplets in solution without the addition of a sensitizer[12] and we suggest that, in the present case, triplet production is likely facilitated by the presence of the heavy Br atoms.

### 2.1.2 Wavepacket analysis

The time-dependent transient absorption kinetics (Fig. 2(d)) show obvious high frequency oscillations at early time delays. These are dynamical signatures of vibrational wavepackets – coherent superpositions of Franck-Condon active vibrations excited by short duration broadband pump pulses.[44–46] The analysis of the vibrational coherences can provide insights into the vibronic activity of the optically bright state and, potentially, of the nuclear motions associated with driving a surface crossing with the optically dark state. Fig. 3(a) shows the first 2 ps of the spectrally resolved chirp-corrected TA spectrum of TDPP-Br. Subtracting the population relaxation dynamics from these data (500 fs $\leq t \leq$ 4000 fs) allows for isolation of the coherent wavepacket dynamics (Fig. 3(b)); note that only times up to 2 ps are shown to highlight oscillations in the false colour contour plot. These data were then fast Fourier-transformed (FFT) along the pump-probe time delay axis to produce a false colour contour map of vibrational wavepacket wavenumber as a function of probe wavelength, shown in Fig. 4(a). Reference to the TA data shown in Fig. 2 encouraged the choice of the 510–560 nm, 600–650 nm and 745–755 nm probe wavelength regions (highlighted by dashed boxes in Fig. 4(a)) to average the vibrational wavepacket dynamics associated with the $1A_g$, $1B_u$ and $2A_g$ electronic states, respectively. The wavelength ranges were carefully chosen to sample as much of the selected transient feature as possible, while minimising spectral overlap with signals originating from other electronic states. These data also contain coherent signatures from the solvent arising from non-resonant impulsive-stimulated Raman scattering (ISRS). To assist in the assignment of these peaks, toluene-only data were acquired under similar experimental conditions and are shown in Fig. S2. The extracted wavenumbers match the known Raman lines of toluene (Table S1).

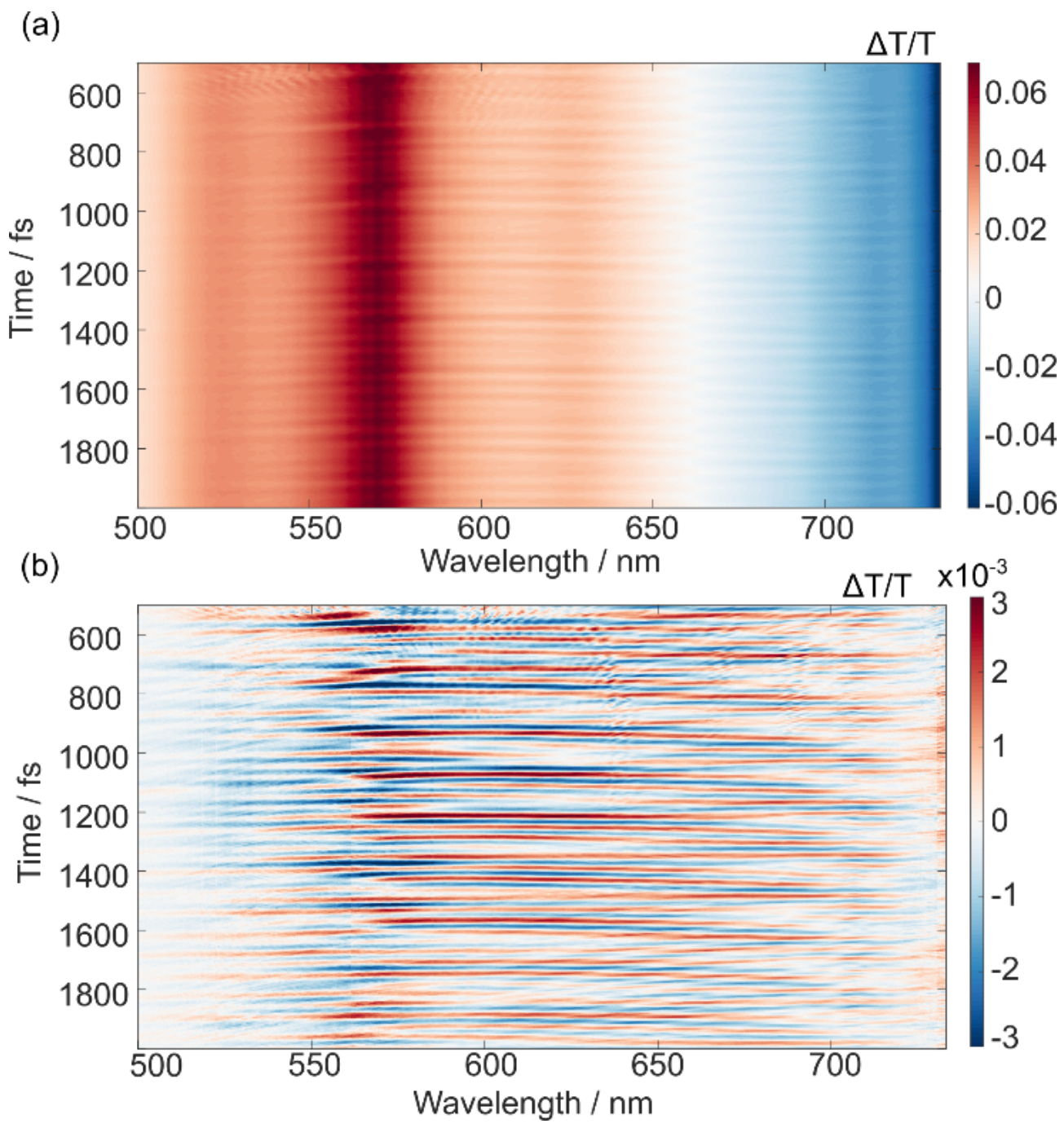

**Fig. 3** Wavepacket dynamics for TDPP-Br dissolved in toluene. False colour contour map of transient absorption data (a) before and (b) after subtraction of electronic population dynamics.

Definitive assignment of the solute modes is very challenging given the high density of vibrational states (the full TDPP-Br molecule has 438 vibrational degrees of freedom). However, DFT/B3LYP/6-311G(d,p) calculations of the ground state normal modes of the core model system using methyl side chains and including both bromine atoms (96 vibrational degrees of freedom) provide many useful insights.

Vibrational wavepackets associated with the TDDP-Br transient GSB were observed at wavenumbers throughout the range 50–1550 $cm^{-1}$ (Fig. 4(a)). For wavenumbers ≤ 700 $cm^{-1}$, the observed features can be plausibly assigned using DFT calculations to in-plane ring breathing and/or in-plane twisting modes involving the thiophene rings and the DPP core, while features at >700 $cm^{-1}$ are consistent with vibrational modes involving C–C/C=C/C–N–C stretching motions along the π-conjugated backbone extending through flanking thiophene units and the DPP core. A full list of the observed wavenumbers and proposed assignments for the GSB wavepackets observed in the TDPP-Br TA data is given in Table S4 of the ESI. All assignments were made by considering calculated wavenumbers within a 5% tolerance of the experimentally observed value. In most instances, this method generated between one and three plausible assignments due to the vast density of vibrational states; however, the associated nuclear motions of the candidate modes were quite similar.

Analysis of the 600–650 nm probe window associated with the $1B_u$ excited state, also returned several vibrational wavepackets in the 250–1550 $cm^{-1}$ range (Fig. 4(b) and listed in Table S5). These features were assigned using vibrational wavenumbers calculated for the smaller TDPP-Br analogue with truncated alkyl sidechains using TD-DFT/B3LYP/6-311G(d,p) calculations of the $1B_u$ minimum energy structure. (B3LYP can be reliably employed in this case due to the minor charge-transfer character of this state.) The feature appearing at 288 $cm^{-1}$ aligns closely with the same peak found in the GSB trace and, again, is likely attributable to in-plane breathing and/or in-plane twisting distortions of the DPP and/or thiophene rings. The appearance of such a feature in both traces is consistent with Franck-Condon activity in the $1B_u \leftarrow 1A_g$ transition. Wavepackets at higher wavenumbers, including one at ~1362 $cm^{-1}$ that matches well with the vibrational peak separation in the TDPP-Br electronic absorption spectrum (Fig. 2(a)), are plausibly assigned to C–C/C=C and/or C–N–C stretching motions. In the trace associated with the $1B_u$ state, much of the vibrational wavepacket activity identified in the 600–650 nm probe window appears to reflect nuclear motions centred on the core DPP unit, consistent with the SF-TD-DFT predicted geometry changes[43] upon photoexcitation (see Figs. S6 and S7 for the reproduced minimum energy structures and associated bond lengths).

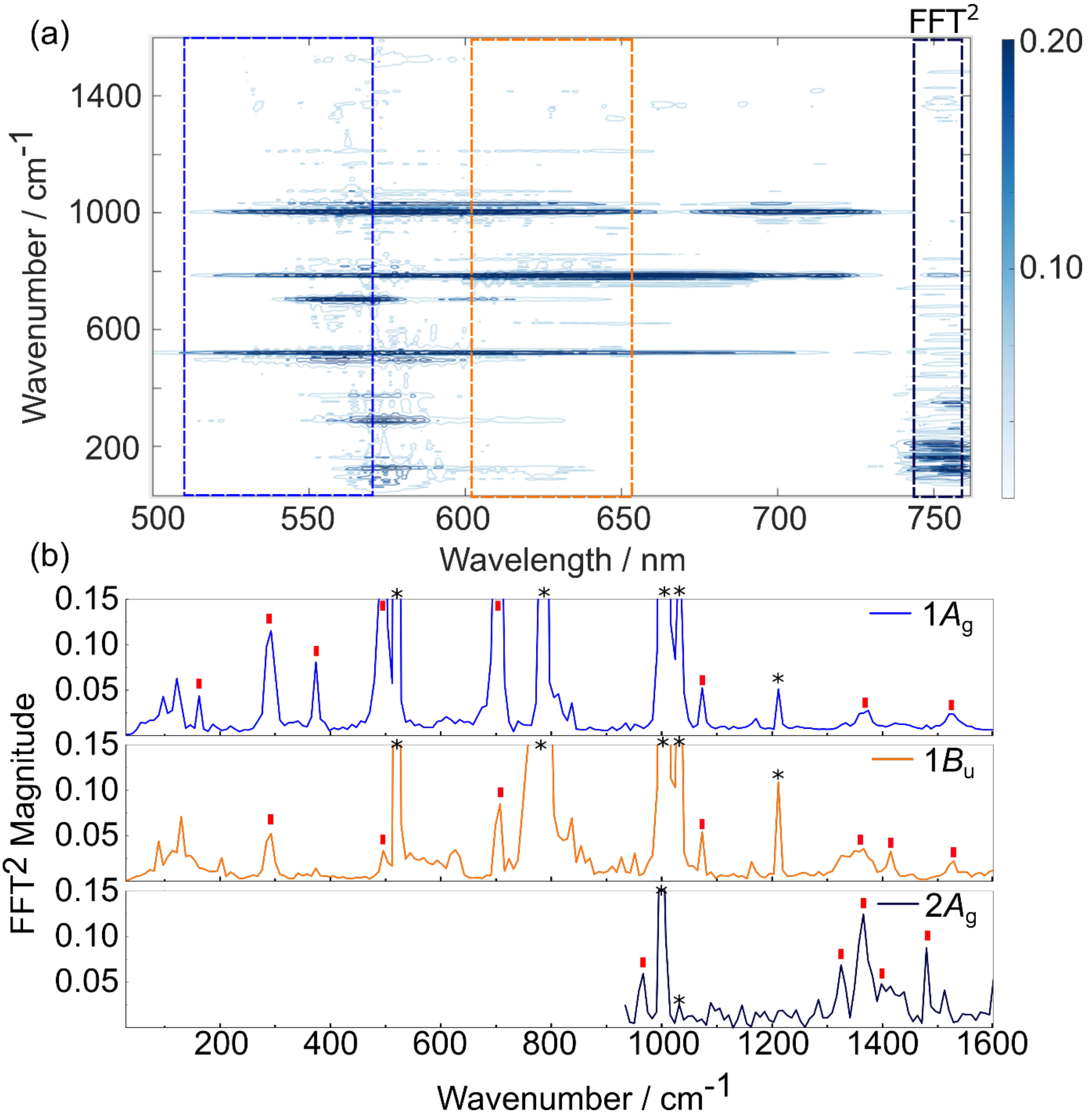

**Fig. 4** TDPP-Br wavepacket dynamics in toluene. (a) Fourier-Transform map of wavepacket-modulated transient absorption spectra of TDDP-Br in toluene. (b) Vibrational wavenumbers associated with the $1A_g$ (510–560 nm), $1B_u$ (600–650 nm), and $2A_g$ (745–755 nm) states. The trace associated with the $2A_g$ state was truncated below 800 cm$^{-1}$ to remove low frequency noise associated with the probe supercontinuum generation. Red ticks in panel (b) indicate assigned TDDP-Br vibrational wavenumbers discussed in the main text and/or in Section 8 of the ESI. Asterisks indicate major peaks associated with toluene solvent.

High frequency vibrational wavepackets were also observed in the 745–755 nm region associated with the $2A_g$ transient. These are more noteworthy, given that the $2A_g \leftarrow 1A_g$ transition has zero-oscillator strength (see Table 1) and thus any population in the $2A_g$ state must arise by ultrafast coupling from (high vibrational levels of) the photoexcited $1B_u$ state. Wavepacket activity in the $2A_g$ state could therefore in principle arise in several ways:

(i) The $1B_u$ and $2A_g$ electronic states have significantly different equilibrium geometries, and if $1B_u \rightarrow 2A_g$ transfer is sufficiently ultrafast, the nascent $2A_g$ molecules would be created far from equilibrium, launching semi-ballistic coherent nuclear motion, reminiscent of the wavepacket dynamics recently observed in a model electron transfer reaction.[47]

(ii) The wavepackets could be created impulsively in the $1B_u$ state by $1B_u \leftarrow 1A_g$ photoexcitation but involve nuclear motions orthogonal to the coordinate that drives the prompt $1B_u \rightarrow 2A_g$ surface crossing. These motions then survive as 'spectators' to the fast coupling, mapping into the corresponding vibrations in the $2A_g$ state, as observed previously in a range of ultrafast non-radiative processes of various molecules;[48–51]

(iii) The wavepackets could reflect nuclear motions that mediate the non-adiabatic transfer in the regions of conical intersection between the states.[52–55] If the crossing between the two states is sufficiently fast, coupling nuclear motions could be created coherently at the intersection and map into $2A_g$ state wavepackets.

Unfortunately, it is not possible to calculate the vibrational normal modes of the $2A_g$ state with SF-TD-DFT- a level of theory required to describe this doubly excited state accurately. Thus, another strategy had to be used to allow insights into the nature of the $2A_g$ state wavepackets. Given the similar calculated minimum energy geometries of the $1B_u$ and $2A_g$ states[43] (*cf.* Figs. S7 and S8 reproduced from Casal *et al.*), we assume that their normal modes of vibration and the associated wavenumbers are also similar. The similar excited state geometries mean that the above scenario (i) is unlikely to be the origin of $2A_g$ wavepackets for TDPP-Br. Scenario (ii) involving wavepackets created upon photoexcitation of the $1B_u$ state and survive the surface crossing to the $2A_g$ state is likely. Indeed, several $2A_g$ peaks can be ascribed to this category: wavenumbers 1365 and 1410 $cm^{-1}$ have credibly assigned motions similar to those identified for the $1B_u$ state. As before, these are plausibly assigned to in-plane C–N–C stretching modes and C–C/C=C stretching motions extending across the DPP and thiophene rings.

Surprisingly, examination of Fig. 4(b) also reveals wavepackets in the $2A_g$ data (at 965, 1325 and 1479 $cm^{-1}$) that have no analogues in the $1B_u$ or $S_0$ transients. As these wavenumbers do not appear in any other trace, the wavepacket motion must have been produced upon $1B_u \rightarrow 2A_g$ interconversion. As such, it is most likely that these reflect nuclear motions created at a $1B_u/2A_g$ conical intersection, *i.e.*, the coupling mode(s) themselves. The overall symmetry of the electronic-nuclear wavefunction must be conserved upon transfer from the $1B_u$ state to the $2A_g$ state, meaning that the coupling nuclear motions must have $b_u$ symmetry (as $B_u \otimes b_u = A_g$), *i.e.* in-plane (preserving the $C_{2h}$ mirror plane symmetry), but anti-symmetric (breaking the inversion symmetry) motions. Consistent with this symmetry requirement, the 965 $cm^{-1}$ wavepacket can be plausibly assigned to a nuclear mode involving anti-symmetric C–S–C stretching on the thiophene units coupled to C–C/C=C stretching motions over the whole π-conjugated system, the 1325 $cm^{-1}$ feature could arise from anti-symmetric DPP and thiophene ring breathing motions, and the 1479 $cm^{-1}$ feature could be assigned to a high amplitude anti-symmetric stretching motion of the central C=C bonds of the DPP unit.

## 2.2 TDPP-v-TDPP

### 2.2.1. Photophysics

The TDPP-v-TDPP absorption in toluene is evidently red-shifted (recall Fig. 1(d)) relative to that of the TDPP-Br monomer, consistent with delocalisation over both the aromatic TDPP units and the vinylene linker, and is in good agreement with the previous report by Mukhopadhyay *et al.*, albeit the prior study used a TDPP-v-TDPP dimer with shorter (and linear) alkyl substituents, and was conducted in chlorobenzene solution.[38] Fig. 5(b) displays TA spectra of TDPP-v-TDPP in toluene for a selection of pump-probe time delays. Three distinct features are evident in the spectra: (i) a positive GSB signal which mirrors the linear absorption spectrum (Fig. 5(a)), (ii) a weak and short-lived ESA centred at ~500 nm, and (iii) a second ESA in the near-IR probe region with a rise time constant of $\tau = 92 \pm 5$ fs. Commensurate with the rise of the near-IR ESA, the 500 nm ESA decays in parallel with a discernible change in GSB line shape on the red-edge, where stimulated emission is anticipated, thus leading to a shift in the isosbestic point at ~710 nm.

Again, comparison with recent DFT/MRCI calculations of a TDPP-v-TDPP variant with truncated alkyl chains[43] proved insightful. For the simplified TDPP-v-TDPP dimer, calculations reveal two excited states at similar excitation energies (see reproduced data from ref. [43] in Table 2): as in TDPP, a bright, primarily singly excited $1B_u$ state and a dark predominantly doubly excited $2A_g$ state. In contrast to TDPP, Casal *et al.* found that the energetic ordering of these two states is reversed: the $2A_g$ state is predicted to be ~0.1 eV lower in energy in the Franck-Condon region and to have an adiabatic excitation energy ~0.5 eV less than that of the $1B_u$ state. The systematic study showed that the $1B_u$/$2A_g$ state ordering is greatly influenced by the length of the π-system;[43] increasing the number of π electrons stabilises both states, but the $2A_g$ state more significantly, in agreement with the inversion of excited states observed experimentally (*vide infra*). By analogy with the preceding discussion of TDPP-Br and theoretical calculations, the 500 nm ESA feature is assigned to the optically bright $1B_u$ state prepared by the initial photoexcitation, which decays rapidly into the lower-energy $2A_g$ state.

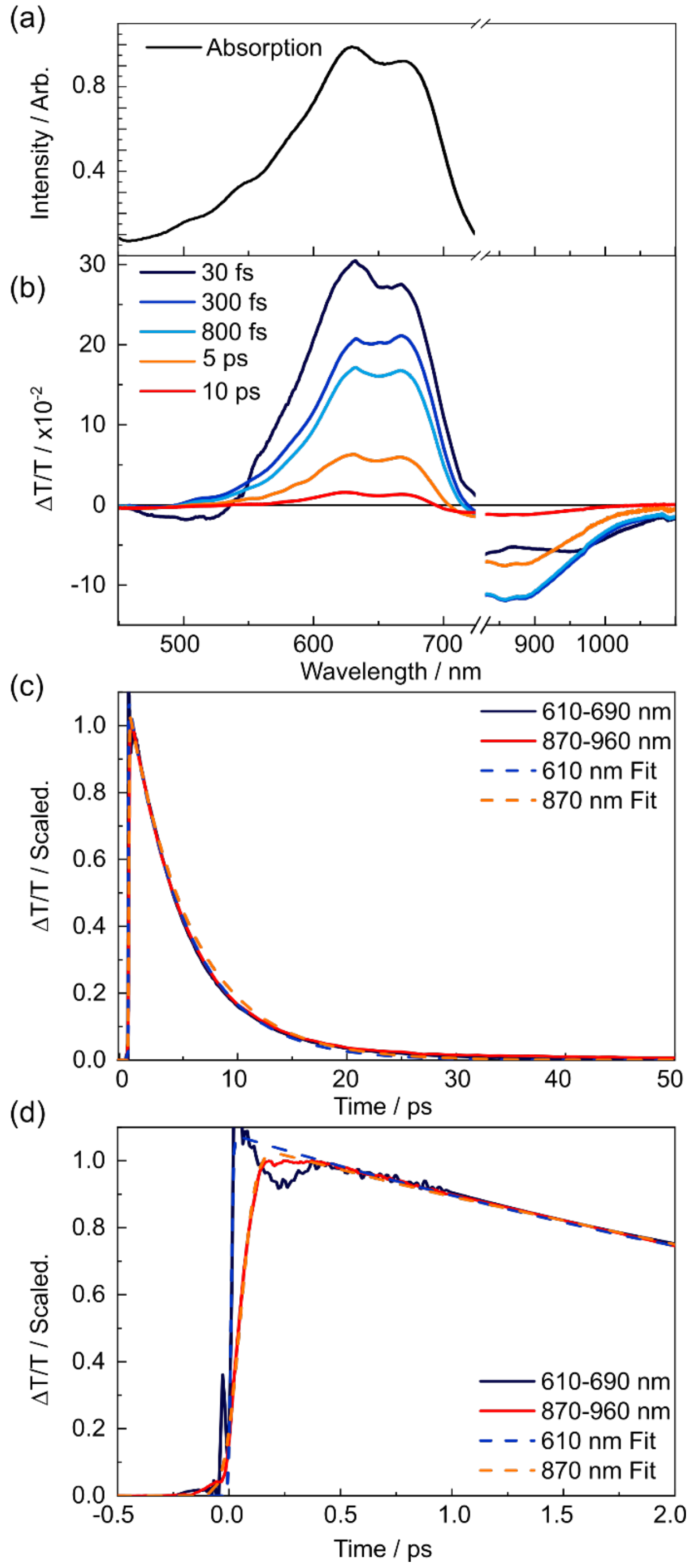

**Fig. 5** Transient absorption spectra and kinetics of TDPP-v-TDPP in toluene solution. (a) Linear absorption spectrum, (b) spectrally resolved TA spectra at selected pump-probe time delays in the range 30 fs ≤ *t* ≤ 10 ps. (c) Scaled wavelength averaged dynamics over the first 50 ps associated with key spectral features and (d) scaled early time dynamics showing the population transfer between the optically bright and lower energy states.

**Table 2** Calculated vertical (VEE) and adiabatic (AEE) excitation energies, oscillator strength ($f$) and percentage of doubly excitation (%D) character for simplified TDPP-v-TDPP. Excitation energies were evaluated at the DFT(BHLYP)/MRCI/SV(P) level. Data are reproduced from ref. [43].

| State | VEE / eV (nm) | AEE / eV (nm) | $S_n \leftarrow S_0$ $f$ | %D ($S_0$ min) |
|---|---|---|---|---|
| $S_1$ ($2A_g$) | 1.57 (790) | 1.23 (1008) | 0.00 | 57.0 |
| $S_2$ ($1B_u$) | 1.66 (747) | 1.46 (849) | 2.10 | 10.4 |

No further spectral evolution was observed at pump-probe time delays beyond 1 ps, and both the GSB and near-IR ESA features were found to decay on the same ~ 5.5 ps timescale (GSB: $\tau = 5.3 \pm 0.4$ ps, ESA: $\tau = 5.7 \pm 0.4$ ps). The kinetics reported here are consistent with a previous TA study of TDPP-v-TDPP in chlorobenzene solution using ~150 fs pump pulses, except that the prior study was unable to fully capture the early time $1B_u/2A_g$ interconversion dynamics.[38]

The early time dynamics have clear parallels with those displayed by the monomer, *i.e.*, very fast coupling between the photoexcited state and the dark doubly excited state. Thereafter the monomer and dimer exhibit strikingly different photophysics. TDPP-Br molecules that remain in the $1B_u$ state survive for nanoseconds in a fluorescent state. Those that couple into the $2A_g$ state display a somewhat shorter lifetime ($\tau$ = 175 ps), determined by the rate of intersystem crossing. In contrast, for TDPP-v-TDPP we observe a full depopulation of the $1B_u$ state, and a much shorter $2A_g$ lifetime of 5.5 ps. The fast and efficient non-radiative decay of the $2A_g$ state necessitates a low-lying, accessible within thermal energies, conical intersection with the $S_0$ state– a pathway seemingly unavailable in monomer TDDP-Br.

### 2.2.2. Wavepacket analysis

A vibrational wavepacket analysis of the first 4 ps of TA data for TDPP-v-TDPP offers further insights into the non-radiative decay of electronically excited dimer molecules. Again, the analysis identifies three wavelength regions: 480–530 nm, 550–600 nm and 850–900 nm, which probe the optically bright ($1B_u$), ground ($1A_g$) and dark ($2A_g$) state populations, respectively (see Fig. 6(a)). The FFT wavepacket-modulated transient absorption false contour plot (Fig. 6(b)) from each region shows peaks attributable to the dimer and solvent ISRS. The studied dimer molecule has a vast number of normal modes (a total of 456 vibrational degrees of freedom), making a definitive assignment of the observed peaks difficult, but comparison with wavenumbers calculated using DFT/B3LYP/6-311G(d,p) and TD-DFT/B3LYP/6-311G(d,p) for the normal modes of TDPP-v-TDPP with truncated alkyl groups allows for possible interpretation.

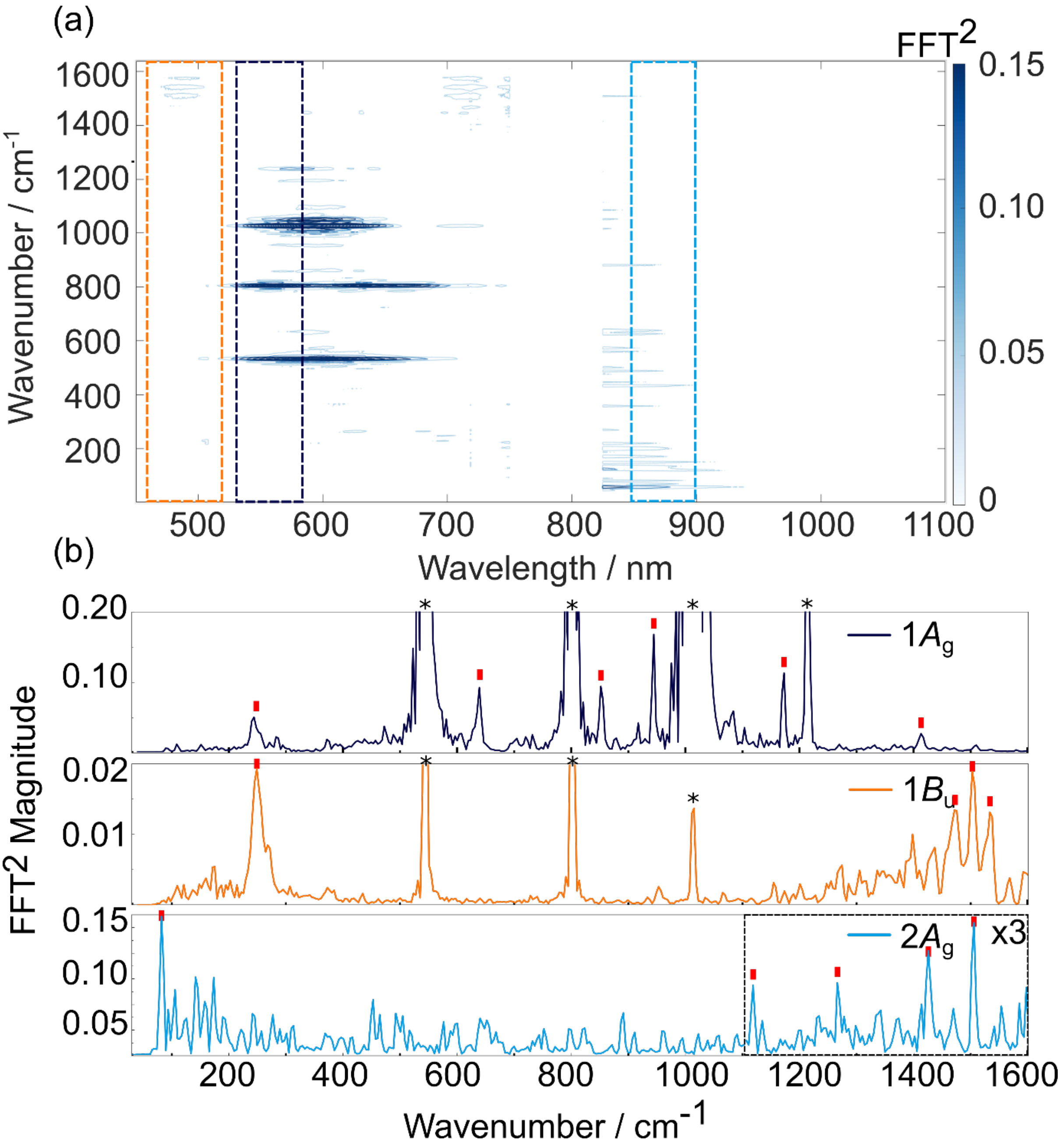

**Fig. 6** Wavepacket dynamics of TDPP-v-TDPP in toluene solution. (a) False colour map of vibrational wavenumber *vs.* probe wavelength extracted from transient absorption measurements. (b) Vibrational wavenumbers associated with the $1A_g$ (550–600 nm), $2A_g$ (850–900 nm) and $1B_u$ (480–530 nm) electronic states. Due to the short lifetime of the $1B_u$ state, the wavepacket intensity is notably weaker. Asterisks indicate solvent vibrations; red labels indicate modes assigned in the main text and in section 8 of the ESI. Asterisks indicate major peaks associated with toluene solvent.

For the $S_0$ and $1B_u$ states, associated wavepacket spectra contain peaks throughout the 200–1550 cm$^{-1}$ range, which can be plausibly assigned to in-plane ring breathing motions and C–C/C=C stretching motions over the whole molecular structure. These assignments are consistent with the predicted geometry changes from the ground to excited state minimum. Full assignment of these wavepackets is given in Tables S7 and S8 of the ESI.

Transient data in the near-IR probe region associated with the $2A_g$ dark intermediate state also exhibits vibrational wavepacket activity and, similar to the TDPP-Br monomer, has two origins. Vibrational activity was observed at 1422 and 1502 cm$^{-1}$, at similar wavenumbers as wavepackets observed in the $1B_u$ or $1A_g$ states, but slightly shifted, suggesting that these wavenumbers correspond to spectator vibrational modes to the $1B_u/2A_g$ surface crossing, and correspond to ring breathing motions of the thiophene and DPP units. At lower wavenumbers three peaks with significant intensity unique to the $2A_g$ state are evident at 54, 1107 and 1258 cm$^{-1}$. As the recent computational study revealed the minimum energy geometries of the $2A_g$ and $1B_u$ states of TDPP-v-TDPP are very similar[43] (associated bond lengths reproduced in Figs. S10 and S11), these new wavepackets must arise from coherent nuclear motion associated with coupling modes created at the $1B_u/2A_g$ conical intersection of the required $b_u$ symmetry which map into $2A_g$ state wavepackets. With these criteria and rationale, it was possible to suggest the following wavepackets assignments: an in-plane anti-symmetric rocking of the DPP units relative to the thiophene rings (with a calculated wavenumber of 50 cm$^{-1}$); DPP/thiophene anti-symmetric ring breathing motions (1113 cm$^{-1}$) and in-plane anti-symmetric C=C stretching motions localised to the central thiophene rings/vinyl linker (1230 cm$^{-1}$).

## 2.3 DPPDTT

The commercially obtained DPPDTT polymer sample had a reported polydispersity index of 2.05 and, unlike TDPP-Br and TDPP-v-TDDP, lacks high order $C_{2h}$ symmetry. The linear absorption spectrum of DPPDTT in chloroform (Fig. 7(a)) is red-shifted relative to that of TDPP-Br, but is similar to that of TDPP-v-TDPP (see comparison in Fig. 1(d)). This similarity hints that the delocalisation length scale in the DPPDTT polymer sample is close to that of the dimer, *i.e.*, 2 repeat units. Fig. 7(b) shows TA spectra for DPPDTT dissolved in chloroform at four representative pump-probe delays, the analysis of which yields the population dynamics reported in this section. Similar TA measurements in toluene (in which the polymer is less soluble) yield poorer signal-to-noise data but return similar kinetics, as described in the ESI (Fig. S3 and S4).

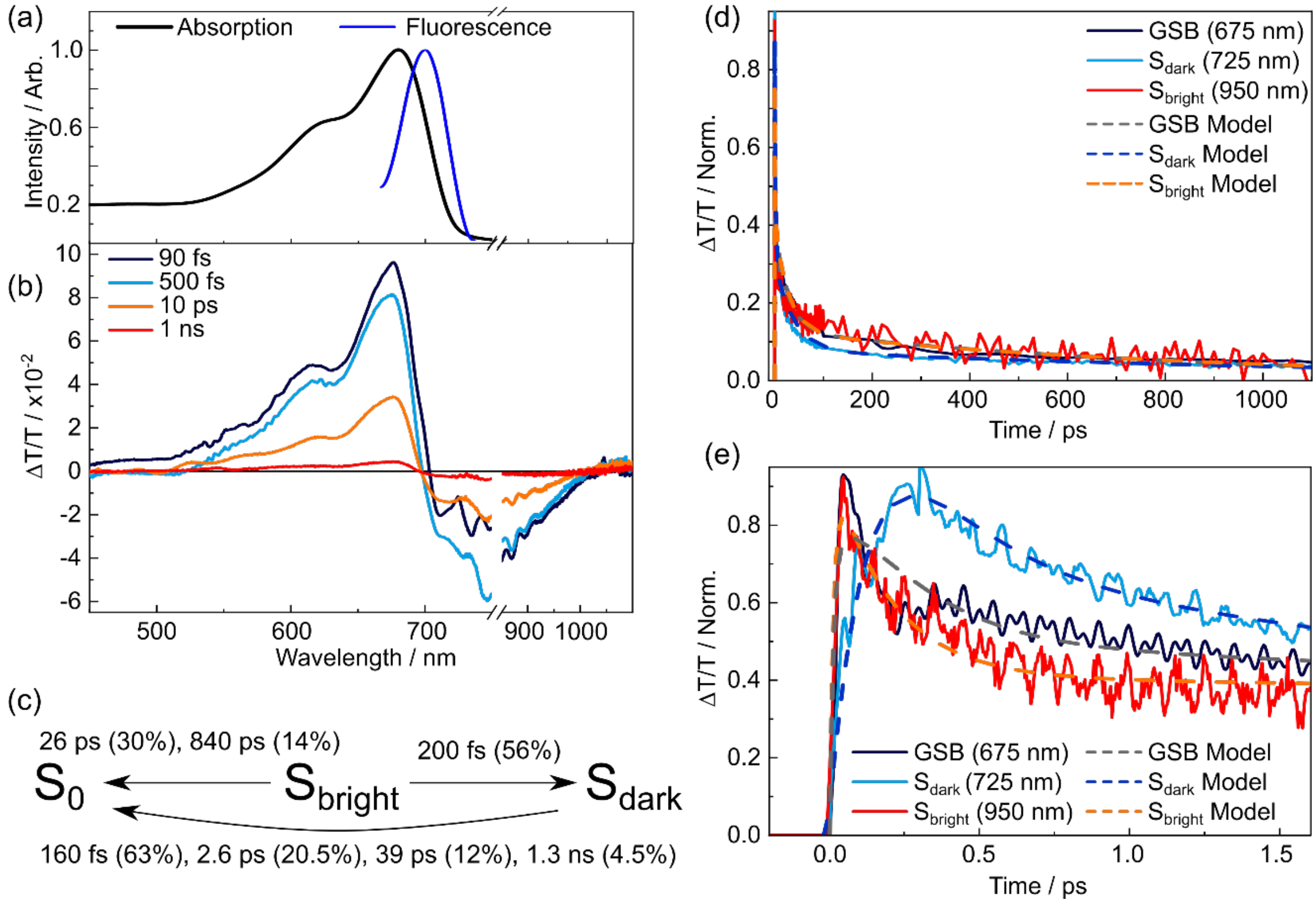

**Fig. 7** Transient absorption spectra and kinetics of DPPDTT in chloroform solution. (a) Linear absorption and fluorescence spectra of DPPDTT in chloroform. (b) Spectrally resolved transient absorption spectra for four different pump-probe time delays. (c) Summary of the rate model used to describe DPPDTT kinetics after excitation to the bright state. Errors associated with lifetimes are reported in the text; the associated amplitudes (given as percentages) are stated in brackets. Normalised kinetics for (d) nanosecond and (e) picosecond dynamics, and include overlaid kinetics returned from the rate model.

The TA spectra for DPPDTT are dominated by two major spectral features, with assignments analogous to the other systems. A large positive signal that spans most of the visible probe window can be assigned to a combination of ground state bleach and stimulated emission, and the negative band peaking in the near-IR to an ESA transient. All these time traces require multiple exponential decay components to fully describe the kinetics, and the ESA feature clearly embodies (at least) two behaviours. Figs. 7(d) and 7(e) compare kinetic data for the two ESAs (725 and 950 nm) and the GSB (675 nm) displayed on two different time axes. The 950 nm ESA rises instantaneously within the instrument response function (IRF), consistent with it being the signature of a directly photoexcited state and thus optically bright in nature (henceforth simply labelled $S_{bright}$). In contrast, the 725 nm ESA rises on a timescale slower than the IRF, consistent with a state formed after the initial photoexcitation. In tandem with the growth of this feature, we note a blue-shift of the isosbestic point at ~700 nm– at probe wavelengths where SE from the bright state would be expected to contribute. The time-dependent shift in the isosbestic point thereby tracks the decay of the fluorescent $S_{bright}$ state. In analogy with TDPP-Br and TDPP-v-TDPP, the slow rise and loss of SE intensity suggest dynamics wherein photoexcitation populates the $S_{bright}$ state (probed by ESA at 950 nm), which rapidly converts to an optically 'dark' state (henceforth labelled $S_{dark}$) that is responsible for the 725 nm ESA.  Despite the 725 nm ESA maximising within ~300 fs, some fraction of the 950 nm ESA signal persists out to beyond 1 ns. Such behaviour has parallels with that observed for the TDPP-Br monomer and suggests incomplete interconversion between the $S_{bright}$ and $S_{dark}$ states, leaving some emissive population in the $S_{bright}$ state.

The observed kinetics are very rich, making fitting individual lifetimes difficult. The TA data at different probe wavelengths could be fit using 3 (or 4) different exponential components, yielding time constants similar to those reported previously for donor-acceptor polymers such as PTB7.[56–58] Such an analysis is difficult to underpin with a physical model but is consistent with polymer samples involving a distribution of chain conformations, but maintaining a delocalisation length of ~2 repeat units, each with different local environments.[56,57] To confirm if a three state model is sufficient to describe the complex kinetics of DPPDTT, a rate model solving a set of coupled differential equations representing the time-dependent populations of the $S_{bright}$, $S_{dark}$ and $S_0$ states was used. These modelled populations are compared to kinetics representing each state to confirm its validity (Fig. 7(d,e)). Full details of the modelling procedure are given in section 2(c) of the ESI.

The initial population of the $S_{bright}$ state is generated by a term which describes the Gaussian IRF. The model allows for several independent populations of the $S_{bright}$ state, each representative of a distribution of polymer conformations in the sample which exhibit different photophysical behaviours. Further, to simplify the model, no exciton diffusion between these domains is allowed on the order of picoseconds, enforcing static inhomogeneity, and thereby allowing the sub-populations to be treated independently. In the model, the $S_{bright}$ state decayed with three-time constants: 200 ± 8 fs (56 %), 26 ± 1 ps (30 %) and 840 ± 40 ps (14 %) where the numbers in parenthesis report the respective normalised amplitudes. The shortest decay component (200 fs) results in formation of the dark state, while the other components of the $S_{bright}$ state decay are associated with direct return to the $S_0$ state *via* radiative or non-radiative mechanisms.

After its formation, $S_{dark}$ decays to $S_0$ as described by four time constants: 160 ± 7 fs (63 %), 2.6 ± 0.1 ps (20.5 %), 39 ± 15 ps (12 %) and 1.3 ± 0.1 ns (4.5 %). The kinetics associated with the GSB are described by an IRF limited rise, and a decay to zero based on a weighted combination of four $S_{dark}$ and two $S_{bright}$ state decay constants. Combining these elements enables prediction of the population relaxation kinetics for the three key states, as summarised in Fig. 7(c). The results returned by the three-state model are overlaid on the experimental data in Figs. 7(d) and 7(e), and a striking agreement between the predicted and experimentally recorded kinetics is found on both picosecond (Fig. 7(d)) and nanosecond (Fig. 7(e)) timescales. The complex kinetics are not unexpected given the heterogeneous nature of the polymer sample, as noted in previous studies of similar systems.[57]

To unravel a physical mechanism underpinning this proposed model and the nature of the $S_{dark}$ state, inspiration was taken from the analysis of other donor–acceptor polymers,[59–62] and guided by our analysis of monomer TDDP-Br. The long wavelength wing of the pump pulses populates low-vibrational levels of the DPPDTT polymer near the $S_{bright}$ minimum. These levels vibrationally relax to the minimum of the $S_{bright}$ potential and then radiatively decay back to $S_0$. The short-lived lifetimes associated with these components (26 and 840 ps), which are similar to those observed for P3HT,[57] hint at non-radiative processes back to $S_0$ that limit the excited state lifetime. This observation explains the weak, yet measurable, fluorescence for DPPDTT as opposed to the non-fluorescent dimer. In analogy with TDDP-Br, DPPDTT polymers excited with the short wavelength edge of the laser pulse, which populates highly vibrationally excited levels of the $S_{bright}$ state, are assumed to follow a similar mechanism to TDDP-Br. The excess vibrational energy within the $S_{bright}$ state, prior to vibrational cooling, allows for interconversion into the $S_{dark}$ state within 200 fs. The $S_{dark}$ state then decays (with $\tau$ = 160 fs, 2.6 ps, 39 ps and 1.3 ns) directly back to $S_0$. Unlike TDPP-Br, however, we hypothesise that the dark state in the asymmetric DPPDTT polymer is charge-transfer, or possibly charge-separated, in character and the myriad of decay rates represent non-radiative charge-recombination processes which depend on the local environment specific to the polymer conformation.

As Fig. 7(d) shows, the DPPDTT transient absorption kinetics also show oscillations attributable to vibrational wavepackets from the polymer along with ISRS from the chloroform solvent. In chloroform, the solvent ISRS signal dwarfs the intensity of solute wavepackets. Therefore, to probe DPPDTT wavepackets, the early time dynamics were measured in toluene solution. The wavepacket analysis for DPPDTT in toluene is reported in Fig. S5 of the ESI. Unlike the monomer and dimer, the size and polydispersity of the polymer sample prevent accurate calculations and thus makes assignment of modes difficult. Wavepacket wavenumbers associated with the $S_{\text{dark}}$ state are noteworthy (618, 734, 859, 946 $cm^{-1}$), and consistent with the deduced ultrafast coupling of $S_{\text{bright}}$ and $S_{\text{dark}}$ states. Three wavepackets (734, 859 and 946 $cm^{-1}$) are evident in both $S_{\text{bright}}$, $S_0$ and $S_{\text{dark}}$ traces, a signature that the surface crossing is prompt and that coherent nuclear motion can carry through the surface crossing into the dark state. A single wavepacket, 618 $cm^{-1}$, however, is only evident in the $S_{\text{dark}}$ trace and therefore is a likely signature of nuclear motions created at the conical intersection linking the two states.

## 2.4 Summary

The energetic ordering of dark ($2A_g$) and bright ($1B_u$) electronic states has a profound influence on the photophysics of the TDPP monomer and dimer. The multireference computational results by Casal *et al.* using SF-TD-DFT and DFT/MRCI[43] were critical to elucidating and characterising the dark $2A_g$ state. This recent computational study highlights the need to use methods which can properly describe double excitations (or to benchmark where appropriate) electronic structure methods, especially when there is no *a priori* knowledge of the nature of the electronically excited states, which may be poorly characterised by standard TD-DFT methods.

To the best of our knowledge, wavepackets originating from coherent nuclear motions associated with coupling modes at conical intersections have only been reported twice before for molecular systems,[54,55] and only one of these studies offered mode assignments based on high-level calculations.[54] The assignment of the wavepacket vibrational modes in the present study was only possible with the use of the corresponding theory and consideration of symmetry requirements. Such an approach could be applied to many existing data in the literature and offer deeper insights into coherent motions in the region of conical intersections. It is important to note that observation of these dynamics in TDPPs is also in part due to three key other fortuitous factors: there is minimal spectral overlap between $2A_g$ and $1B_u$ ESA transients, which greatly simplifies spectral assignments and means wavepacket signatures can be readily associated with a single electronic state; the minimal spectral shifting present in the TA data, presumably due to the small/zero structural re-organisation between the $2A_g$ and $1B_u$ states; the prompt surface crossings ensured vibrational dephasing/intra-molecular vibrational distribution is incomplete by the time the $1B_u/2A_g$ conical intersection is encountered and that the decoherence of nuclear motion is sufficiently minimal that new coherences created at the conical intersection map into observable $2A_g$ wavepackets.

## 3. Conclusions

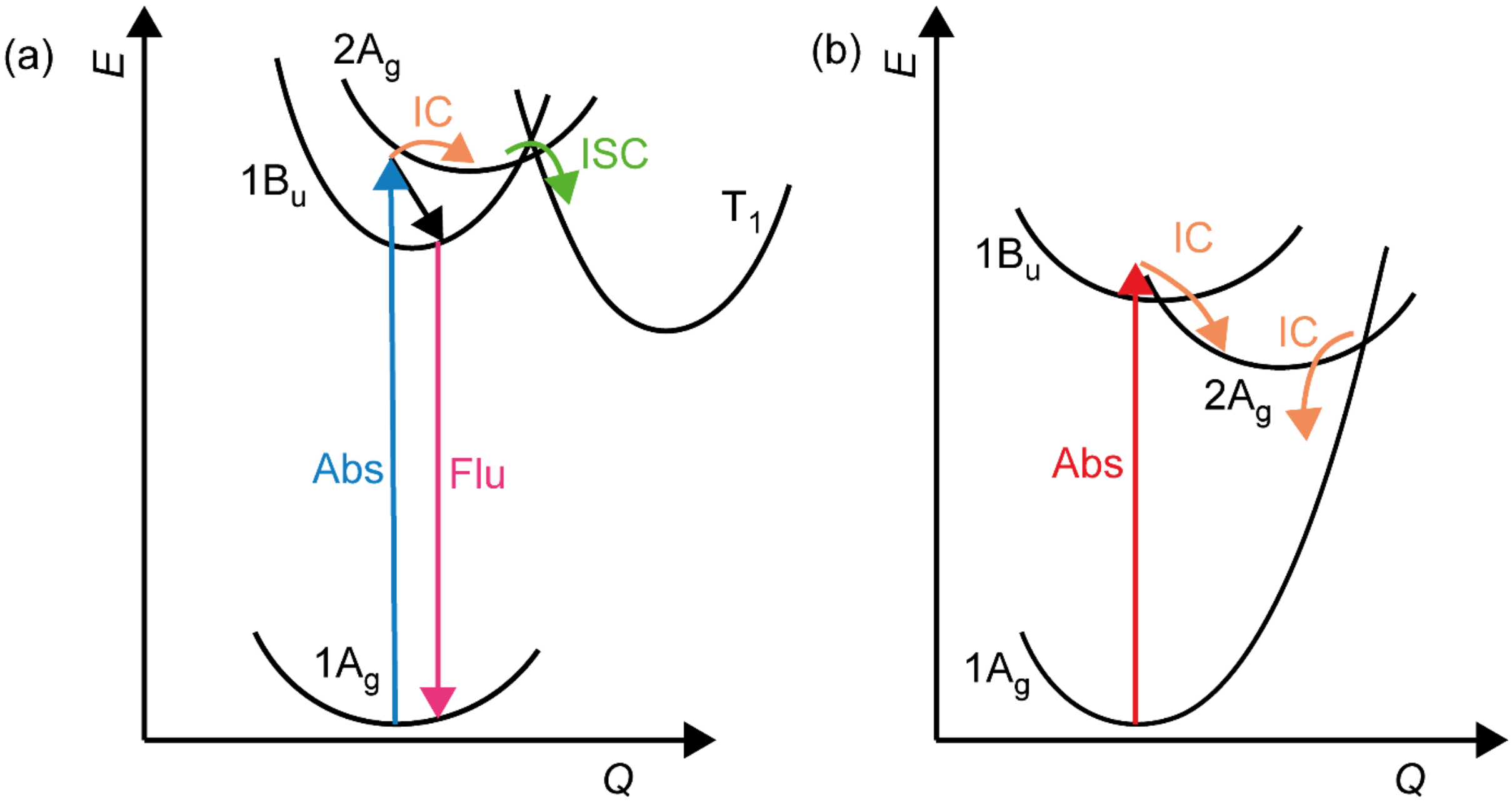

**Fig. 8** Schematic potential energy curves and illustrative pathways representing the major excited state dynamics for (a) TDPP-Br, and (b) TDPP-v-TDPP.

The photophysics of TDPP-Br, TDPP-v-TDPP and DPPDTT upon broadband visible excitation are dictated by two electronically excited states. Photoexcitation populates a dipole-allowed excited state, which in the $C_{2h}$ point group is of $1B_u$ symmetry for monomer TDPP-Br and the TDPP-v-TDPP dimer. In the monomer, the $1B_u$ state is the lowest energy singlet excited state, with a ~ 5 ns lifetime. A small fraction of the $1B_u$ population that is created in vibrationally excited levels, are able to convert into a different electronic state ($2A_g$), which has significant doubly excited character.[43] In TDPP-Br, the $2A_g$ state acted as a precursor for rapid intersystem crossing into a triplet state (Fig 8(a)). In the dimer, the dark $2A_g$ state lies below the $1B_u$ state in the vertical Franck-Condon region. Analysis of the transient kinetics shows that the vast majority of excited molecules are converted into the lower-lying $2A_g$ state within <

100 fs and subsequently couple back to the $S_0$ state on a tens of picosecond timescale, with high efficiency (Fig 8(b)).

The DPPDTT polymer exhibits some dynamical and electronic structural similarities with the monomer and dimer. The delocalisation length scale of the polymer sample is deduced to be similar to that of the dimer, spanning ~2 repeat units. However, the excited state dynamics and kinetics of the polymer more closely resemble those of TDPP-Br, but with two notable differences. The interconversion of the optically bright state into the dark state is more efficient in DPPDTT than the monomer, and the electronic character of the dark state in the polymer is charge-transfer in nature, supporting subsequent electron/hole transport within polymer chains. Furthermore, due to the inhomogeneity of the polymer sample, each dynamical process exhibits a range of lifetimes originating from differences in the surrounding environment.

The ultrashort laser pulses used in these studies inevitably launched excited state wavepacket dynamics. Wavepackets were observed in the bright $1B_u$ state and, surprisingly, also the optically dark $2A_g$ state. These wavepackets are deduced to be formed via two different mechanisms: (i) wavepackets generated upon photoexcitation of the $1B_u$ state but have associated nuclear motions that lie orthogonal to the $1B_u/2A_g$ crossing, meaning the vibrational coherence is unaffected by the crossing and transferred into the second electronic state. This phenomenon has been observed previously in other ultrafast condensed phase studies involving fast non-radiative transfer between two electronic states on a timescale faster than the associated vibrational dephasing.[48–51] (ii) More remarkable, is the observation of *new wavepackets* unique to the optically dark $2A_g$ state, born from coherent nuclear motions generated at the $1B_u/2A_g$ conical intersection. These motions, the so-called coupling modes, are created out of necessity to couple the electronic state wavefunctions and map into $2A_g$ vibrational wavepackets. This phenomenon has only been reported twice before in molecular condensed phase studies.[54,55]

Our study emphasises how the excited state dynamics of a family of prototypical TDPP molecules is finely tuned by the relative energetic ordering of two electronic states of very different character. The present experimental study, along with recent computational results[43], also highlights the need for caution when invoking models to explain the excited state phenomena of TDPPs, without strong experimental and/or theoretical evidence. Finally, the wavepacket dynamics observed here further illustrate the exquisite detail that can be extracted about the non-radiative relaxation dynamics in condensed phase studies of modest sized molecular systems.

## 4. Methods

### 4.1 Sample preparation

TDPP-Br (3,6-Bis(5-bromo-2-thienyl)-2,5-bis(2-octyldodecyl)-2,5-dihydropyrrolo[3,4-c]pyrrole-1,4-dione, >98%) and DPPDTT (poly[2,5-(2-octyldodecyl)-3,6-diketopyrrolopyrrole-alt-5,5-(2,5-di(thien-2-yl)thieno [3,2-b]thiophene)], PDI: 2.03) were purchased from Osilla and used without any further purification (DPPDTT is also known as *p*DPPT-TT). TDPP-v-TDPP (*E*)-6,6'-(ethene-1,2-diylbis(thiophene-5,2-diyl))bis(2,5-bis(2-ethylhexyl)-3-(thiophen-2-yl)-2,5-dihydropyrrolo[3,4-*c*]pyrrole-1,4-dione) was synthesized by the Pd-catalysed Stille cross-coupling of bis(tributylstannyl)ethene with mono-brominated TDPP, following the methodology of Patil *et al.*[63] The reaction afforded a mix of the desired product and a major by-product, which were separated by a mixture of column chromatography over silica followed by preparative recycling gel permeation chromatography. TDPP-v-TDPP was characterized by $^{1}$H NMR, $^{13}$C NMR and mass spectra (see Figs. S14-19). The by-product was identified as the homo-coupled TDPP product (TDPP-TDPP) based on NMR and mass spectroscopy measurements, full details of the synthesis and purification procedures are given in section 6 of the ESI.

All molecules were dissolved in either toluene or chloroform depending on their solubility. Solvents were purchased from Sigma Aldrich at HPLC grade (99.8% purity). For TA measurements, the solutions were diluted to have an absorbance between ~0.3–0.4 at the peak wavelength of the pump laser in static 1 mm pathlength cuvettes. Fresh solutions were prepared before each measurement, and no sign of degradation was observed in visible absorption spectra acquired before and after ultrafast spectroscopic measurements. The fluorescence quantum yield of TDDP-Br was determined using a commercial fluorimeter and integrating sphere (Edinburgh Instruments, FS5).

### 4.2 Ultrafast transient absorption spectroscopy

A 1 kHz, 1W, 800 nm commercial laser system (Coherent, Libra) was used to pump a homebuilt non-collinear optical parametric amplifier (NOPA) to produce the pump pulses. The homebuilt NOPA was tuned to produce broadband pump pulses centred at 580 nm with a ~75 nm FWHM (Fig. S1). Pump pulses were compressed using chirped mirrors (Layertec) and fine-tuned with a pair of fused-silica wedges (FemtoOptics, Newport). The pulse duration was determined using polarisation gated frequency resolved optical gating to be 22 fs (FWHM) and limited by third order dispersion (Fig. S1). The compressed pump pulses were attenuated to below 100 nJ using an achromatic polariser–waveplate pair and focused into a ~150 μm spot size at the sample with a 25 cm focal length curved mirror. A very small portion of the Libra output was used to generate white light supercontinuum probe pulses in sapphire or yttrium aluminium garnet crystals, producing probes spanning the visible and near-IR wavelength regions. The pump–probe time delay was controlled by a mechanical delay stage (Physik Instrumente, M-521.DD1). After the sample, the probe and collinear signal was collimated and focused into a Czerny-Turner spectrograph (Shamrock 163, Andor) coupled to a linear CCD array detector (Entwicklungsbüro Stresing) and digitised. All transient measurements were conducted under the magic angle condition. Data were chirp corrected by fitting the solvent only transient response to a high order polynomial.

### 4.3 Density Functional Theory Calculations

DFT and TD-DFT calculations were performed using the Gaussian 16 computational suite[64] to investigate the effect of alkyl group length on the ground state geometry and to calculate ground and excited bright state normal modes and associated wavenumbers. These calculations used the B3LYP exchange-correlation function and a 6-311G(d,p) basis set. A full list of vibrational assignments of wavepacket data are given in section 7 of the ESI.

## Author Contributions

D.W.P., G.A. and O.P. acquired ultrafast and steady state experimental data. M.T.d. C. and J.M.T analysed the data in the light of DFT/MRCI and SF-TD-DFT calculations. X.H. synthesised, purified and characterised TDPP-v-TDPP. D.W.P. analysed all time-resolved data and wrote the first and subsequent major drafts of the manuscript. M.H., M.B., M.N.R.A and T.A.A.O. supervised the project. All authors contributed to the writing of the manuscript and approved the final version.

## Conflicts of interest

The authors have no conflicts to declare.

## Acknowledgements

The authors at Bristol and Aix-Marseille acknowledge funding from the European Union's Horizon 2020 Research and Innovation programme, under grant agreement 828753 (Boostcrop). T.A.A.O. acknowledges financial support from the Royal Society for a University Research Fellowship (UF1402310 and URF\R\201007) and Research Fellows Enhancement Awards (RGF\EA\180076 and RF\ERE\210045). The authors at Aix-Marseille Université acknowledge the Centre de Calcul Intensif d'Aix-Marseille for granting access to its high-performance computing resources and the HPC/AI resources from GENCI-TGCC (Grant 2022-A0110813035). M.H. thanks the Royal Society and the Wolfson Foundation for financial support.

# Supporting information for:
# Probing the electronic structure and photophysics of thiophene–diketopyrrolopyrrole derivatives in solution

Daniel W. Polak,[1] Mariana T. do Casal,[2] Josene M. Toldo,[2] Xiantao Hu,[3]
Giordano Amoruso,[1] Olivia Pomeranc,[1] Martin Heeney,[3]
Mario Barbatti,[2,4] Michael N. R. Ashfold,[1] and Thomas A. A. Oliver[1]*

[1] School of Chemistry, Cantock's Close, University of Bristol, Bristol, BS8 1TS, UK

[2] Aix-Marseille Université, CNRS, ICR, Marseille, France

[3] Department of Chemistry and Centre for Processable Electronics, Imperial College London, White City Campus, London, W12 0BZ, UK

[4] Institut Universitaire de France, 75231, Paris, France

Author for correspondence: tom.oliver@bristol.ac.uk

# 1. Laser Pump Pulse Characterisation

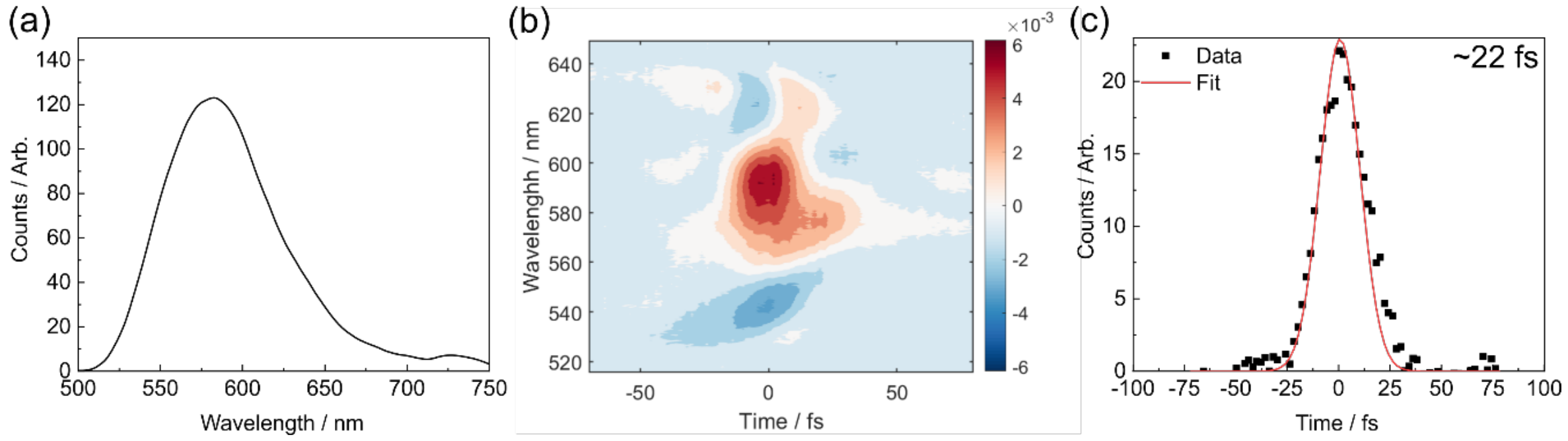

**Fig. S1** (a) Pump laser spectrum used in TA measurements, (b) Polarised-gating frequency resolved optical gating (PG-FROG) characterisation of pump pulses, (c) autocorrelation trace extracted from PG-FROG spectrum (square points) overlaid with Gaussian fit to data (red solid line).

## 2. Transient Absorption Spectroscopy

### (a) Wavepacket analysis of Toluene

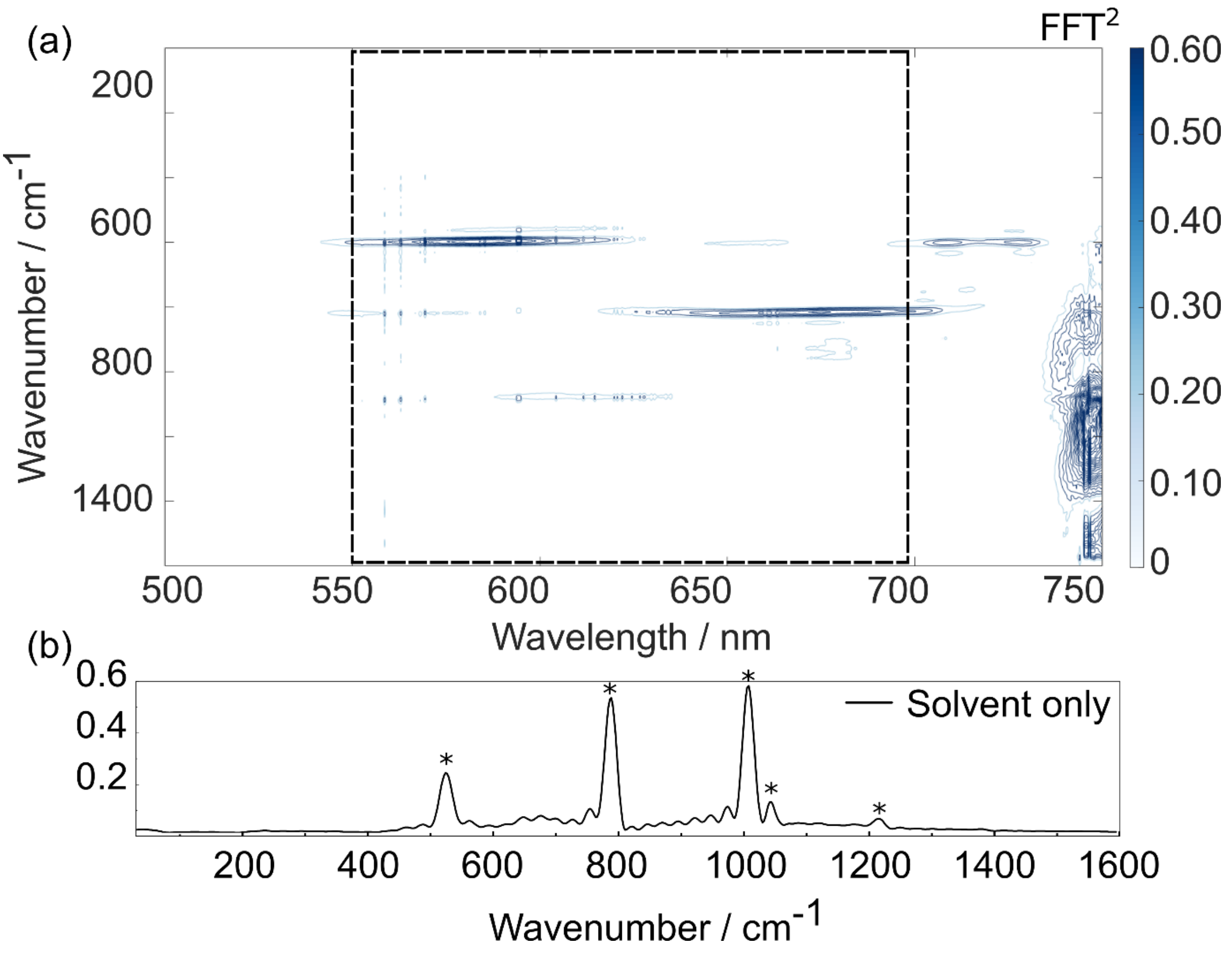

**Fig. S2** Toluene non-resonant impulse stimulated Raman scattering (ISRS) wavepacket dynamics. (a) False colour map of vibrational wavepacket wavenumbers as a function of probe wavelength. (b) Vibrational wavenumbers associated with toluene averaged over the 550–700 nm window. A list of the observed wavenumbers associated with toluene signals is given in Table S1.

**Table S1** Wavenumbers associated with peaks in the averaged toluene spectrum shown in Fig. S2 compared to previously experimentally determined Raman wavenumbers of toluene (from refs [1,2]).

| Observed Wavenumber / $cm^{-1}$ | Literature value / $cm^{-1}$ |
|---|---|
| 523 | 522 |
| 784 | 787 |
| 1000 | 1004 |
| 1032 | 1031 |
| 1208 | 1211 |

## (b) DPPDTT in Toluene

The main manuscript details transient measurements of DPPDTT acquired in chloroform solution. The equivalent data acquired in toluene are displayed in Fig. S3(a). Due to the low solubility of the polymer and some precipitation in toluene, significant scattering is observed in the associated TA data, resulting in far lower signal-to-noise ratios compared to data obtained in chloroform (Fig. 7). To alleviate some of these issues, the frequency and time domains were smoothed by adjacent averaging and the use of a Savitzky-Golay filter, respectively. A comparison of the kinetics for DPPDTT in the two solvents for three probe wavelengths are shown in Fig. S3(b-d), and show clear similarities.

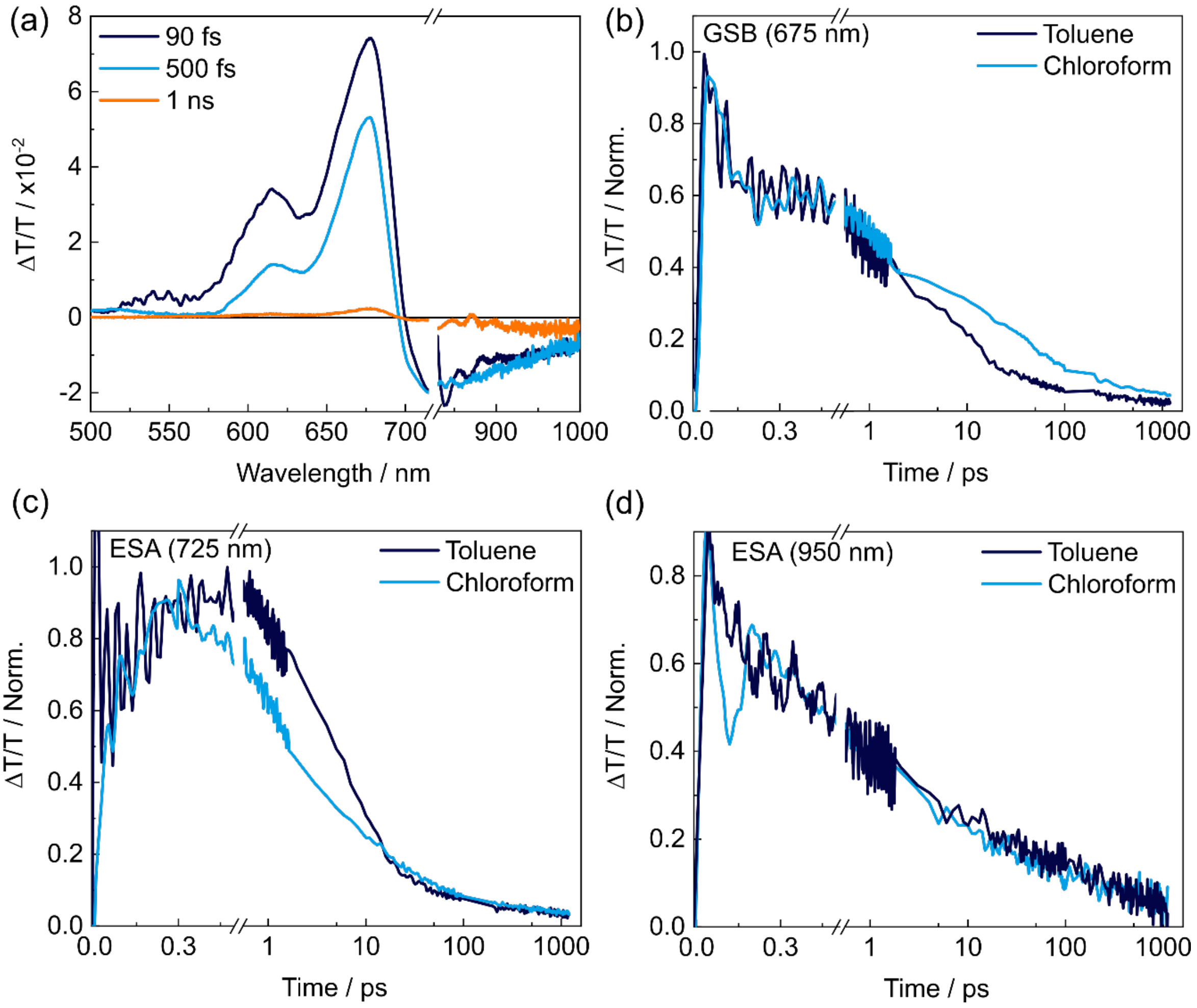

**Fig. S3** (a) DPPDTT transient absorption spectra in toluene solution and (b-d) comparison of kinetics in toluene and chloroform for three probe wavelengths. Toluene kinetics have been smoothed via application of a Savitzky-Golay filter.

### (c) DPPDTT Rate Model

A rate model for both chloroform and toluene measurements was constructed to dissect the complex polymer dynamics: the photoexcited kinetics of DPPDTT were described using a generic rate model (eq. S1), where $y(n)$ and $y(m)$ are the population of the $n^{th}$ and $m^{th}$ states. $k_m$ is the decay rate constant for $y(m)$, which also describes the rate of $y(n)$ formation. The population of $y(n)$ subsequently decays with rate constant $k_n$. Combined, the time dependent population of $y(n)$ can be described:

$$\frac{dy(n)}{dt} = k_m y(m) - k_n y(n) \,. \qquad \text{(S1)}$$

The model includes three electronic states: The ground state ($S_0$), a bright state which is optically prepared ($S_{\text{bright}}$) and a dark intermediate state ($S_{\text{dark}}$). $S_{\text{bright}}$ is generated with a generation function, $G$, that describes the IRF and also serves to depopulate $S_0$. $S_{\text{bright}}$ can subsequently convert into $S_{\text{dark}}$, and both states subsequently decay in parallel back to $S_0$. The model assumes that there are several independent populations of each state generated which do not interconvert on the picosecond time scale as justified in the text below. The constructed equations are then solved numerically using a Matlab script.

There are two possible origins for the range of lifetimes observed in the polymer, arising from either large-scale conformational inhomogeneity, such as planar *vs.* twisted polymer chains, or differences in conjugation lengths on different chain segments. As discussed in the main text, because the absorption maximum of the polymer matches that of the dimer, the delocalisation is considered to be spread over ~2 repeat units (Fig. 1(d))- a conclusion supported by a recent computational study of DPPDTT.[3] Further, if a range of effective conjugation lengths were present, significant broadening of the absorption spectrum would be expected and the evident vibronic progression would not be well resolved. Together this suggests the observed lifetimes represent different sub-ensemble populations corresponding to different polymer geometries (all affording a delocalisation length scale of ~2 repeat units) each with a unique environment. These conformations are static on the timescale of the experiment but have differing photophysical properties leading to the range of reported lifetimes.

The kinetic fits to chloroform data are displayed/discussed in the main text (Fig. 7). A similar model was constructed for DPPDTT in toluene, and the results from this analysis are displayed in Fig. S4. Notably some of the time constants are altered, which is proposed to be due to slight changes in the relative energies of the electronic states, however no additional kinetic components are required to describe the data. Given the excellent match between experimental data and modelled kinetics, it is evident that a three-state model robustly describes the excited state dynamics of DPPDTT.

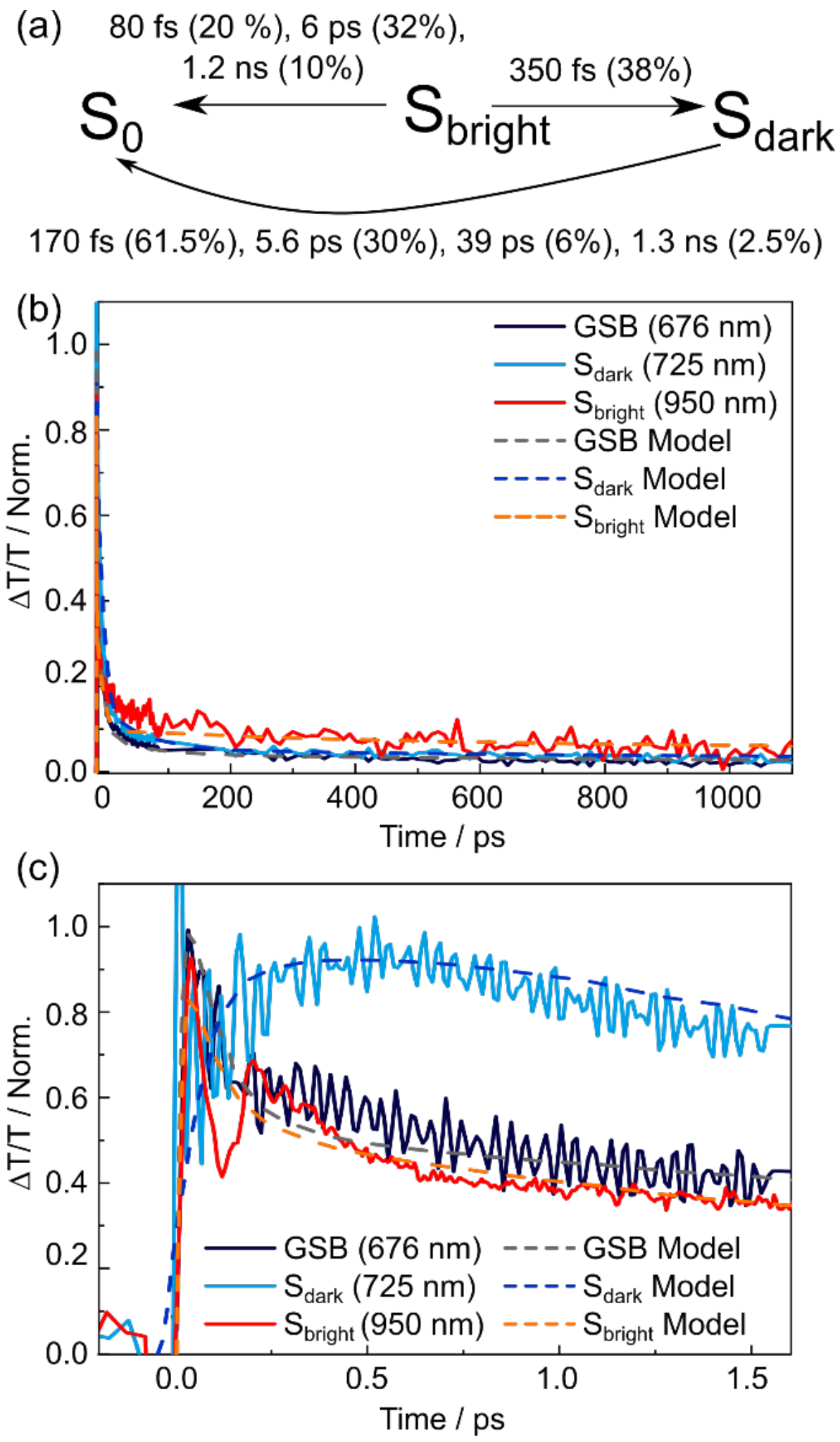

**Fig. S4** (a) Summary of the rate model used to describe DPPDTT kinetics in toluene solution. The associated amplitudes with each lifetime component (as percentages) are given in brackets, the amplitudes have been normalised to match the GSB kinetics. Normalised kinetics for the (b) nanosecond and (c) picosecond dynamic regimes, with overlaid fits (dashed lines) returned from rate model.

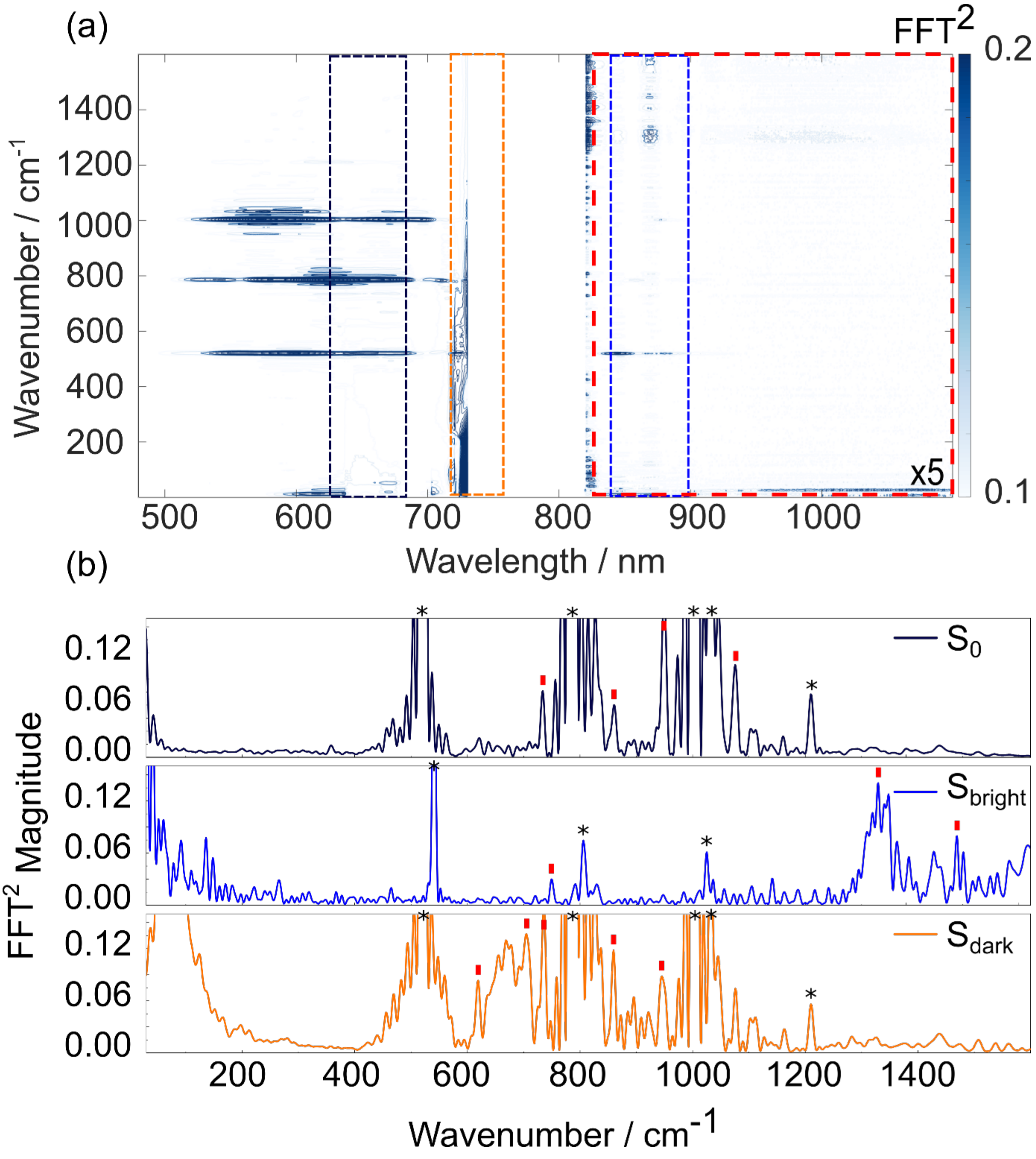

**Fig. S5** Wavepacket dynamics of DPPDTT in toluene solution. (a) False colour map of vibrational wavenumber *vs.* probe wavelength extracted from transient absorption data, (b) vibrational wavenumbers associated with the GSB (625–675 nm), and near-IR ESAs (710–730 nm and 850–900 nm). Asterisks indicate solvent vibrations; red tick labels indicate assigned polymer modes discussed in section 8 of the ESI.

## 3. Kinetic Fitting Parameters

All data were chirp corrected prior to kinetic analysis. The fits to data shown in the main manuscript were modelled using an analytical expression comprised from the convolution of a gaussian IRF with multiple exponential rise/decays. The instrument response function was locked to the pump pulse duration determined by PG-FROG.

**Table S2** Kinetic fitting parameters for TDPP-Br and TDPP-v-TDPP studied in toluene solution.

| **Molecule** | **Probe Wavelength** | **Rise / fs** | **$\tau_1$ / fs (Amp/%)** | **$\tau_2$ / ps (Amp/%)** | **$\tau_3$ / ns (Amp/%)** | **$R^2$** |
|---|---|---|---|---|---|---|
| TDPP-Br | 615 nm (SE) | IRF limited | 103 ± 30 (59) | — | 5 ± 1 (41) | 0.92 |
| TDPP-Br | 750 nm (ESA) | 105 ± 5 | — | 175 ± 5 (73) | 5 ± 1 ns (27) | 0.99 |
| TDPP-v-TDPP | 610 nm (GSB) | IRF limited | — | 5.3 ± 0.4 (100) | — | 0.98 |
| TDPP-v-TDPP | 870 nm (ESA) | 92 ± 5 | — | 5.7 ± 0.4 (100) | — | 0.99 |

## 4. Electronic Structure Calculations

### (a) TDPP

Minimum energy structures for $S_0$ ($1A_g$), $1B_u$ and $2A_g$ states of simplified TDPP molecule are reproduced from Casal *et al.*.[4]

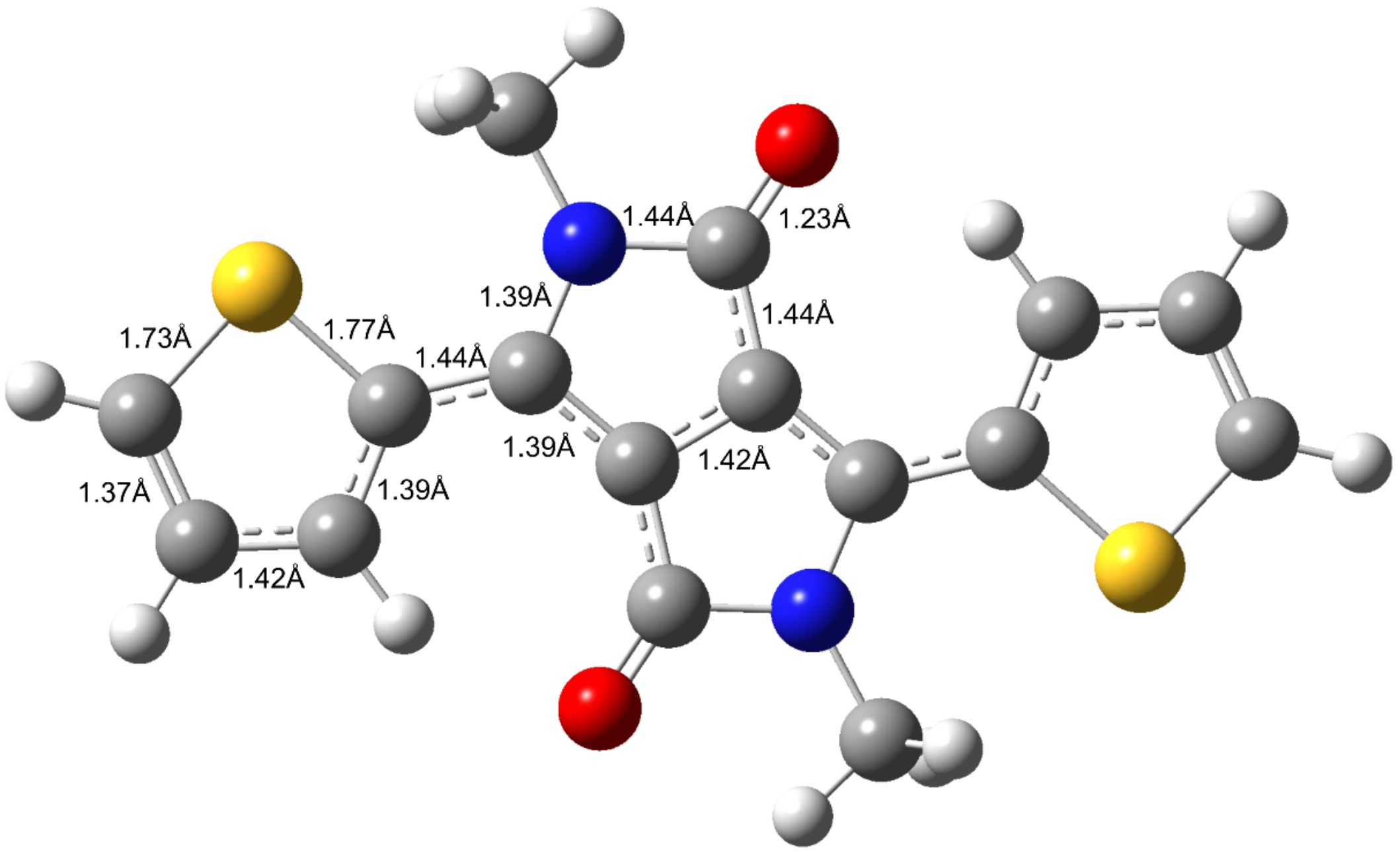

**Fig. S6** Minimum energy geometry and associated bond lengths for $S_0$ ($1A_g$) state of TDPP calculated at the DFT/B3LYP-D3/6-31G(d,p) level of theory.

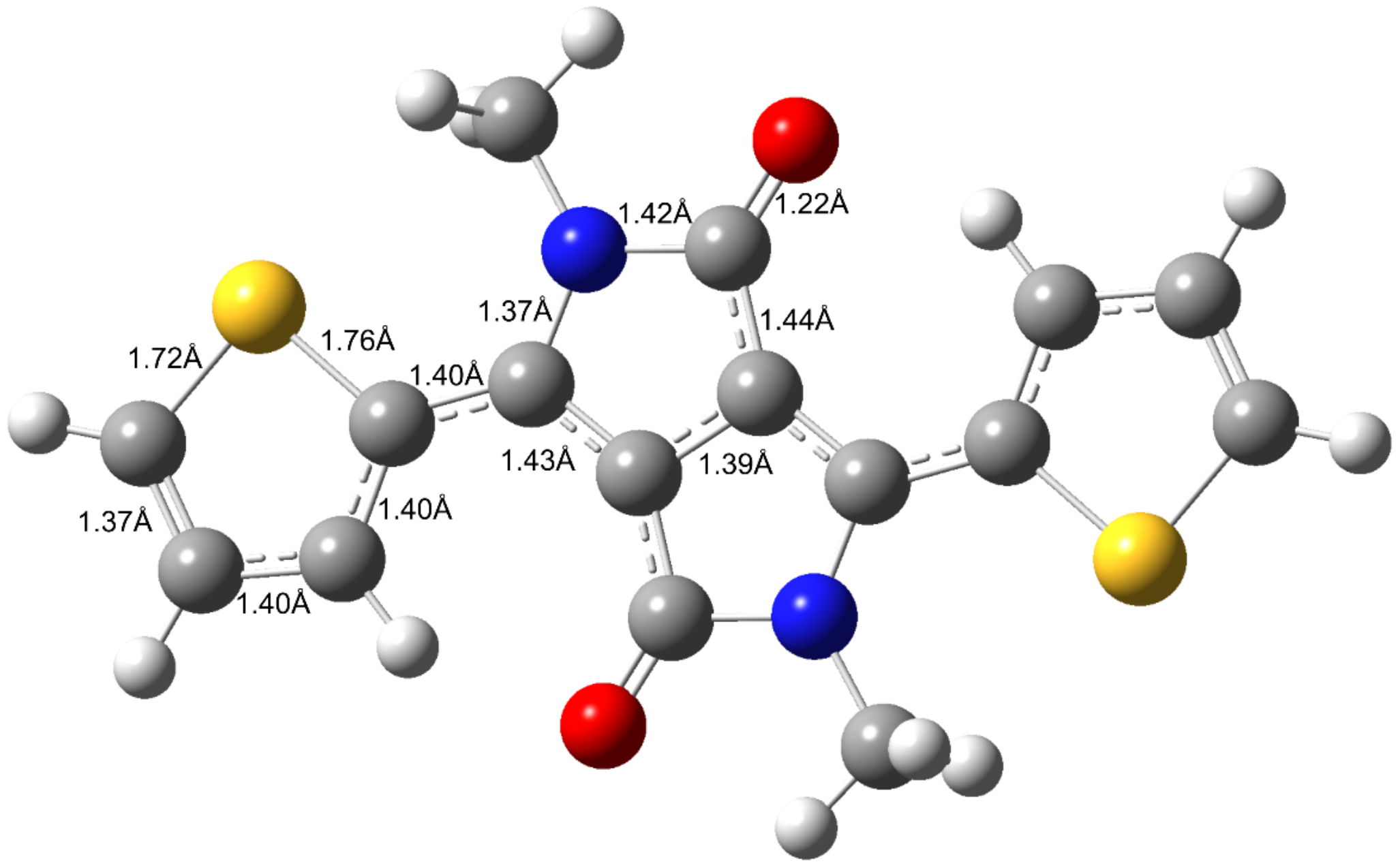

**Fig. S7** Minimum energy geometry and associated bond lengths for $S_1$ ($1B_u$) state of TDPP calculated at the SF-TD-DFT(BHHLYP)/6-31G(d,p) level of theory.

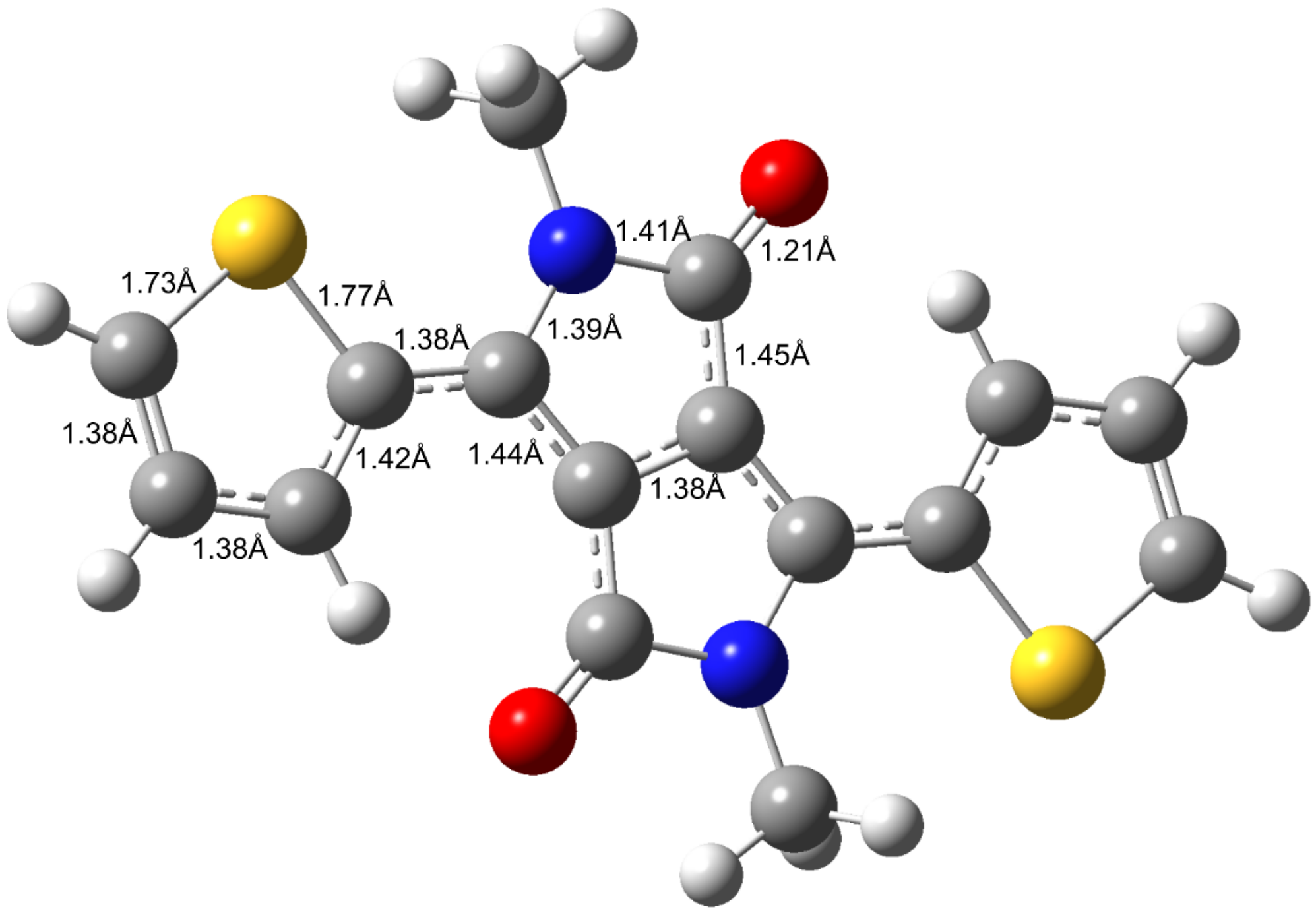

**Fig. S8** Minimum energy geometry and associated bond lengths for $S_2$ ($2A_g$) state of TDPP calculated at the SF-TD-DFT(BHHLYP)/6-31G(d,p) level of theory.

## (b) TDPP-v-TDPP

Minimum energy structures for $S_0$ ($1A_g$), $1B_u$ and $2A_g$ states of simplified TDPP-v-TDPP molecule are reproduced from Casal *et al.*.[4]

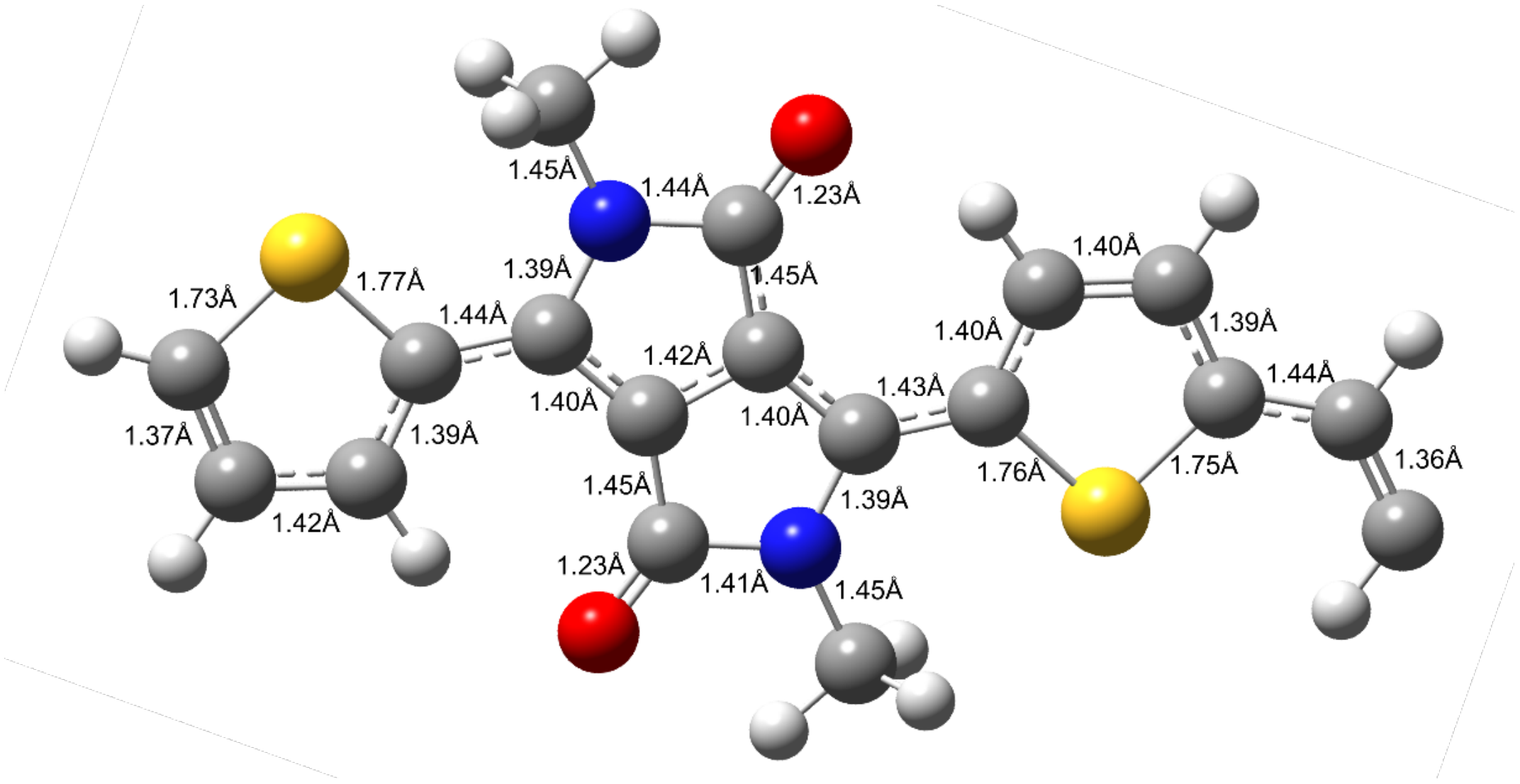

**Fig. S9** Minimum energy geometry and associated bond lengths for $S_0$ ($1A_g$) state of TDPP-v-TDPP calculated at the SF-TD-DFT(BHHLYP)/6-31G(d,p) with enforced $C_{2h}$ symmetry. Only half of the symmetric structure is shown for sake of clarity.

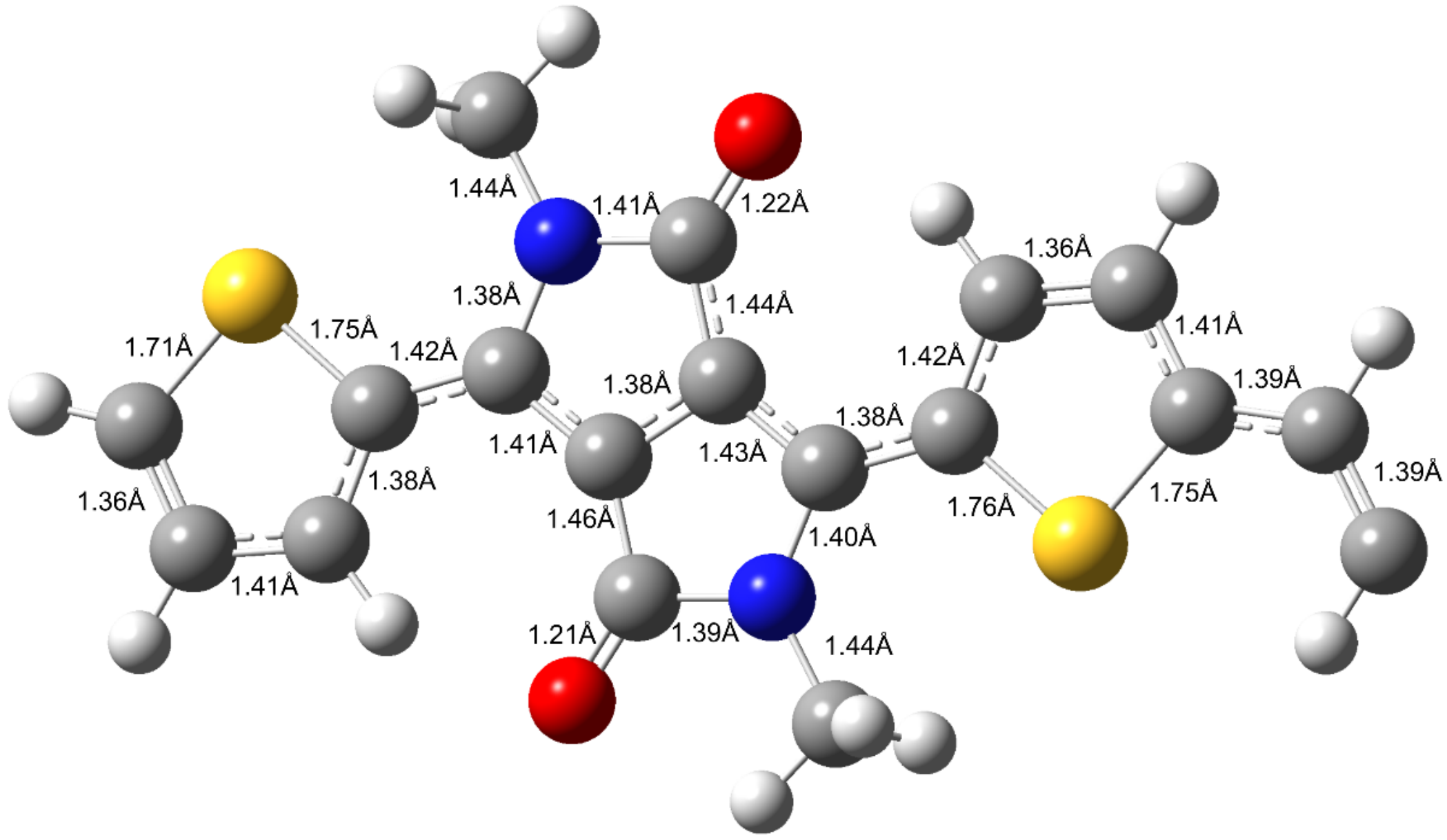

**Fig. S10** Minimum energy geometry and associated bond lengths for $S_1$ ($2A_g$) state of TDPP-v-TDPP calculated at the SF-TD-DFT(BHHLYP)/6-31G(d,p) level of theory with enforced $C_{2h}$ symmetry. Only half of the symmetric structure is shown for sake of clarity.

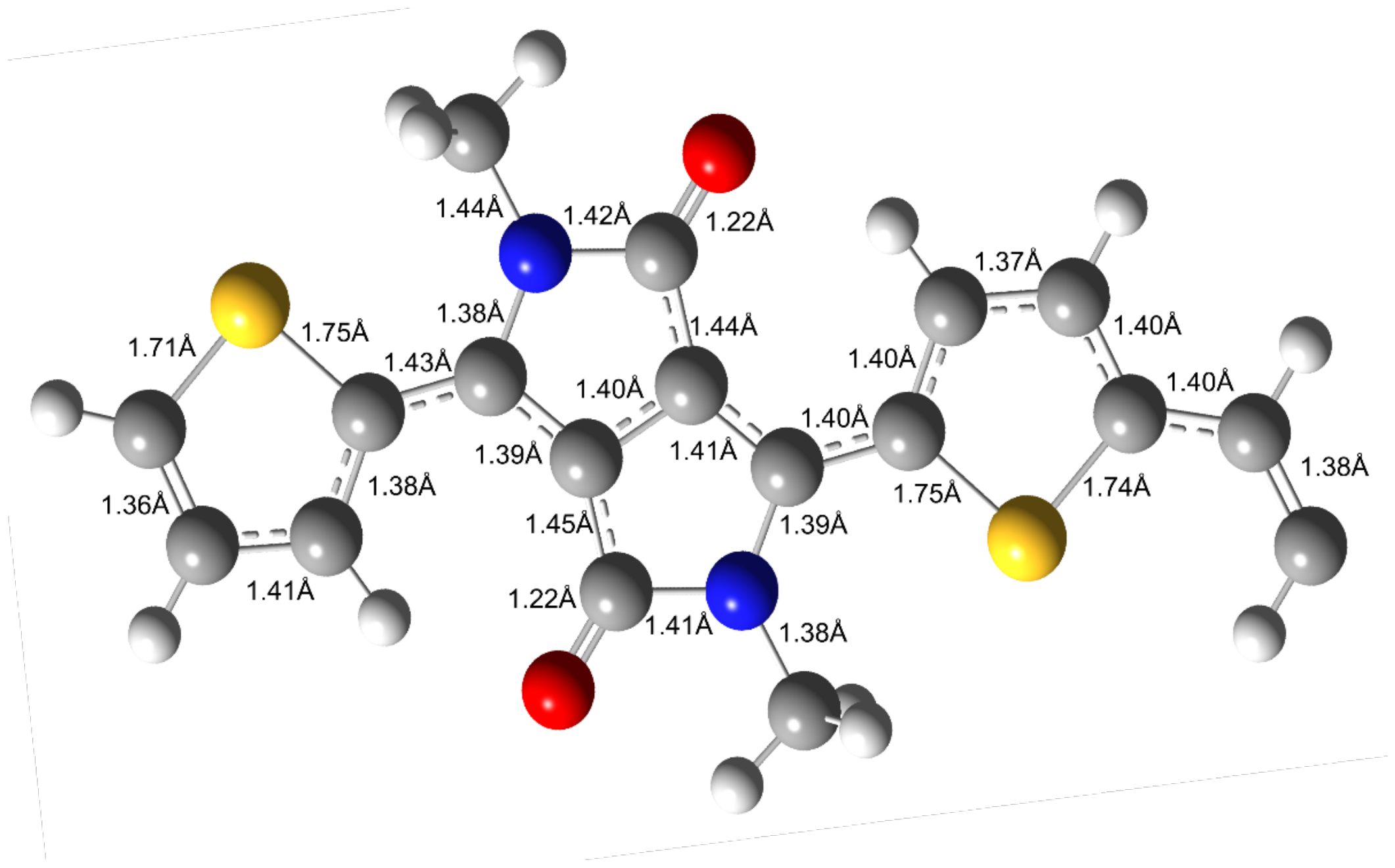

**Fig. S11** Minimum energy geometry and associated bond lengths for $S_2$ ($1B_u$) state of TDPP-v-TDPP calculated at the SF-TD-DFT(BHHLYP)/6-31G(d,p) level of theory with enforced $C_{2h}$ symmetry. Only half of the symmetric structure is shown for sake of clarity.

# 6. TDPP-v-TDPP and TDPP synthetic methods

Commercially available reagents were purchased and used without further purification unless otherwise stated. 3-(5-Bromothiophen-2-yl)-2,5-bis(2-ethylhexyl)-6-(thiophen-2-yl)-2,5-dihydropyrrolo[3,4-c]pyrrole-1,4-dione was synthesized following the published route.[5] $^{1}H$ NMR and $^{13}C$ NMR spectra were recorded on Bruker AV-400 spectrometer in $CDCl_3$ solvent at room temperature. Matrix-assisted laser desorption/ionisation time-of-flight (MALDI-TOF) mass spectrometry was performed on a Bruker ultrafleXtreme MALDI-TOF analyser.

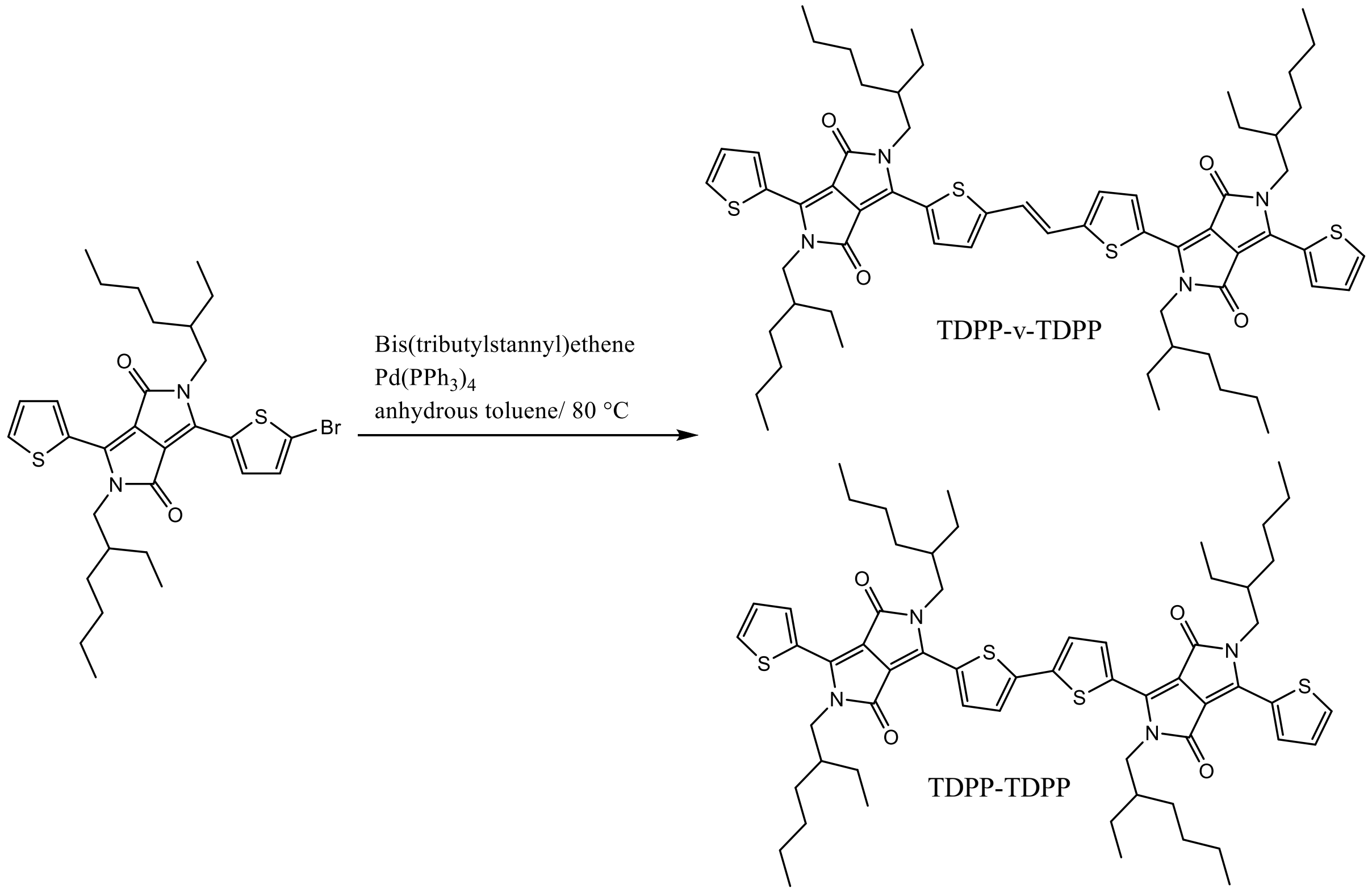

**Scheme 1.** Synthetic dimerization producing TDPP-v-TDPP and TDPP-TDPP.

Compound 3-(5-bromothiophen-2-yl)-2,5-bis(2-ethylhexyl)-6-(thiophen-2-yl)-2,5-dihydropyrrolo[3,4-c]pyrrole-1,4-dione (0.18 g, 0.30 mmol) and bis(tributylstannyl)ethene (0.08 ml, 0.15 mmol) were dissolved in anhydrous toluene (20 ml) in a 100 ml round bottom flask under nitrogen atmosphere. $Pd(PPh_3)_4$ (0.069 g, 0.060 mmol) was added and the mixture stirred at 80 °C for 16 h. After cooling to room temperature, the organic solvent was removed under vacuum and the crude product was purified by column chromatography over silica (eluent: hexane:dichloromethane (1:1)), followed by preparative recycling GPC using chloroform as the eluent. The product and major by product were isolated as dark purple solids (TDPP-v-TDPP: 0.071 g, yield: 44%; TDPP-TDPP: 0.083 g, yield: 53%).

**TDPP-v-TDPP** $^{1}$H NMR (400 MHz, $CD_2Cl_2$) δ 8.92 (2H, d, *J* = 4.11 Hz), 8.88 (2H, d, *J* = 4.13 Hz), 7.68 (2H, d, *J* = 5.02 Hz), 7.30 (4H, t, *J* = 4.23 Hz), 7.27 (2H, s), 4.03 (4H, m), 1.87 (2H, m), 1.31 (16H, m), 0.88 (12H, m). $^{13}$C NMR (101 MHz, $CDCl_3$) δ 161.79, 161.69, 146.97, 140.31, 139.61, 136.54, 135.57, 130.74, 129.99, 129.44, 128.59, 128.45, 122.86, 108.86, 108.41, 46.10, 39.37, 39.22, 30.46, 30.37, 28.70, 28.50, 23.80, 23.70, 23.23, 23.19, 14.24, 14.13, 10.67, 10.62. MS (m/z): [M+] calcd. for C62H80N4O4S4: 1073.59, Found: 1073.0 (MALDI-TOF).

**TDPP-TDPP** $^{1}$H NMR (400 MHz, $CD_2Cl_2$) δ 8.92 (2H, d, *J* = 3.89 Hz), 8.87 (2H, d, *J* = 4.16 Hz), 7.71 (2H, d, *J* = 5.07 Hz), 7.47 (2H, d, *J* = 4.19 Hz), 7.30 (2H, t, *J* = 4.49 Hz), 4.01 (4H, m), 1.85 (2H, m), 1.34 (16H, m), 0.88 (12H, m). $^{13}$C NMR (101 MHz, $CDCl_3$) δ 161.73, 161.58, 141.11, 138.91, 135.92, 135.26, 133.64, 131.82, 131.11, 129.86, 128.63, 126.94, 109.20, 108.17, 90.53, 46.19, 46.09, 39.28, 39.19, 30.35, 30.28, 28.48, 23.68, 23.17, 14.15, 14.12, 10.61, 10.58. MS (m/z): [M+] calcd. for C60H78N4O4S4: 1047.55, Found: 1070.8 (MALDI-TOF).

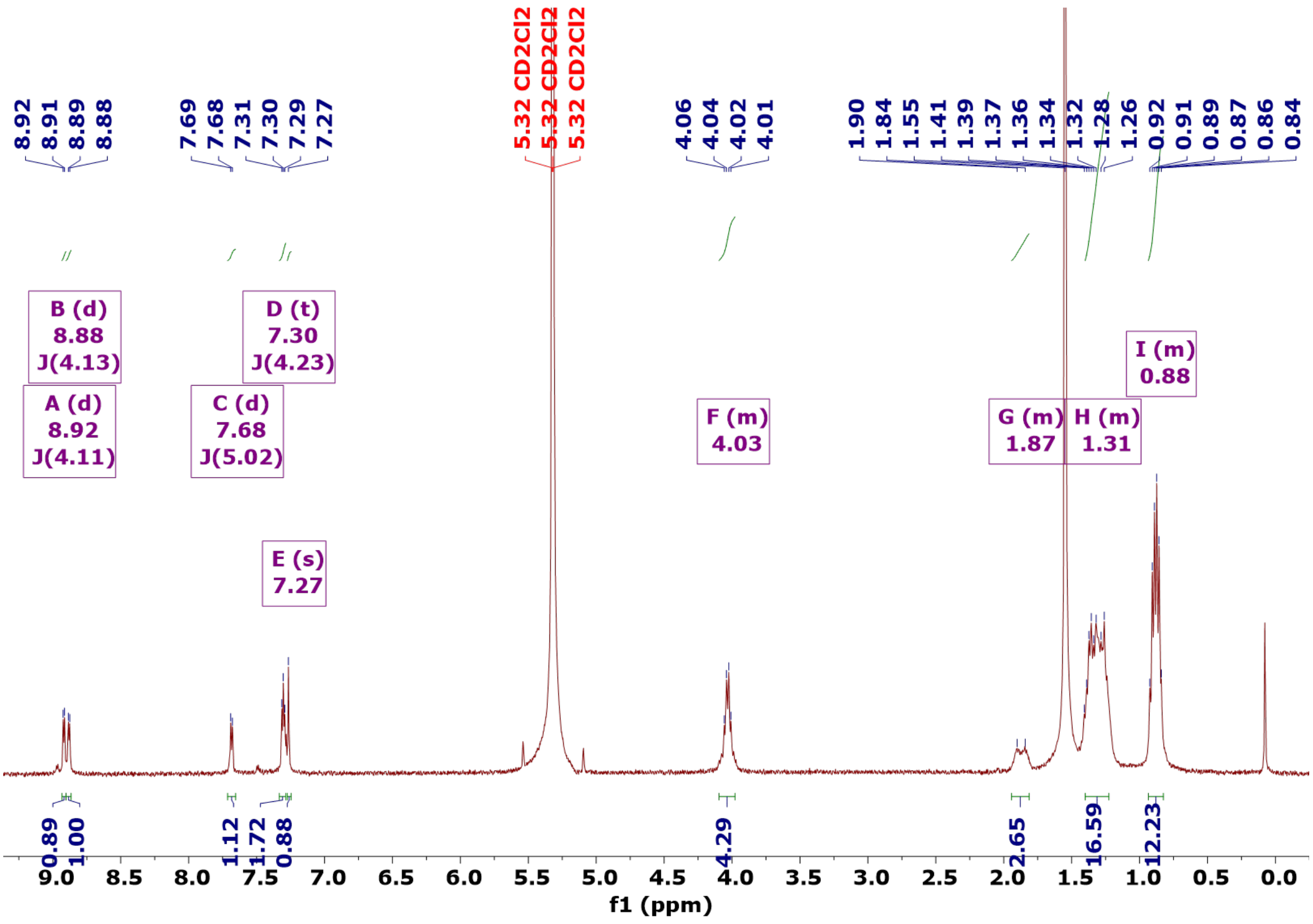

**Fig. S12** $^{1}$H NMR spectrum of TDPP-v-TDPP

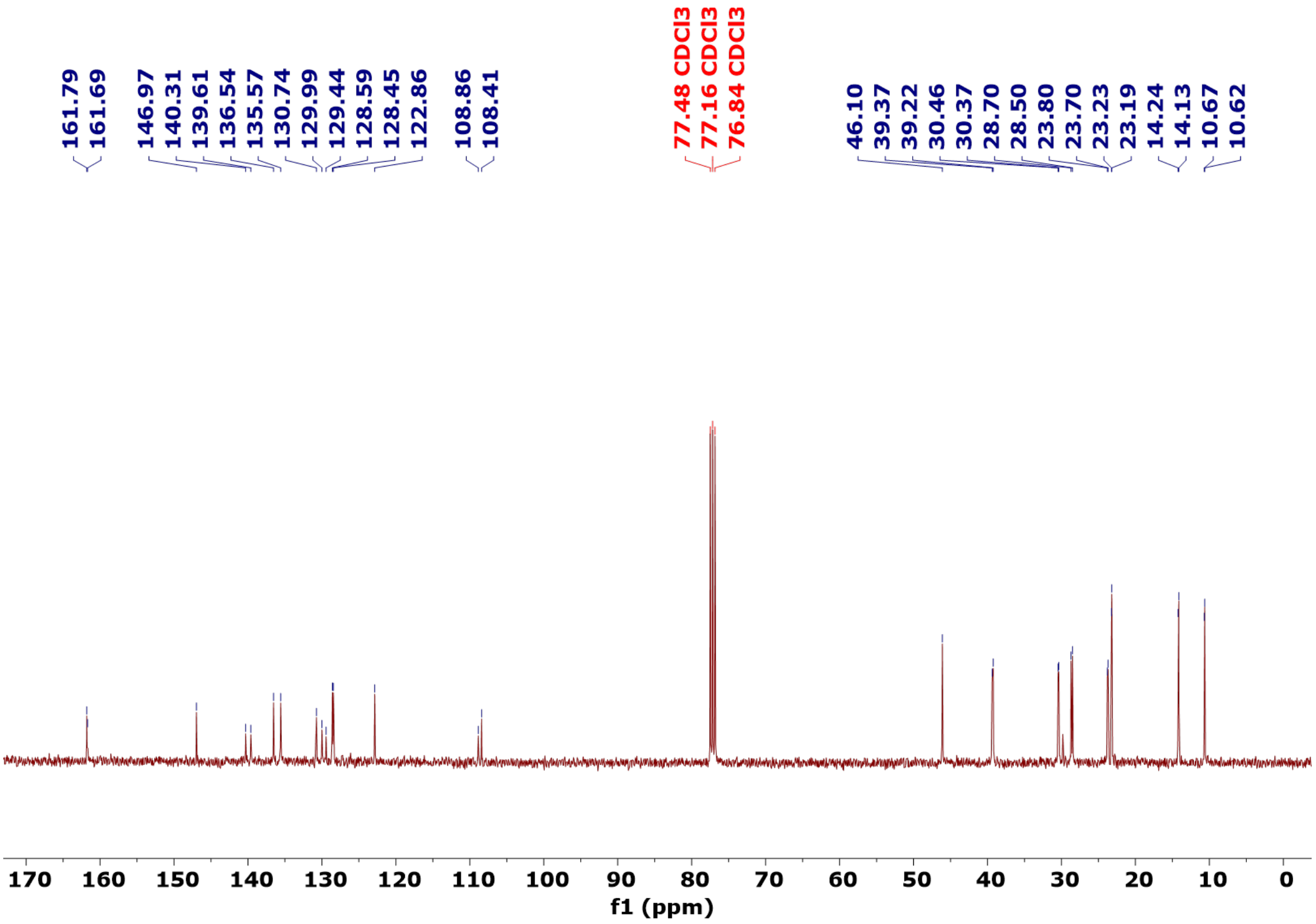

**Fig. S13** $^{13}$C NMR spectrum of TDPP-v-TDPP

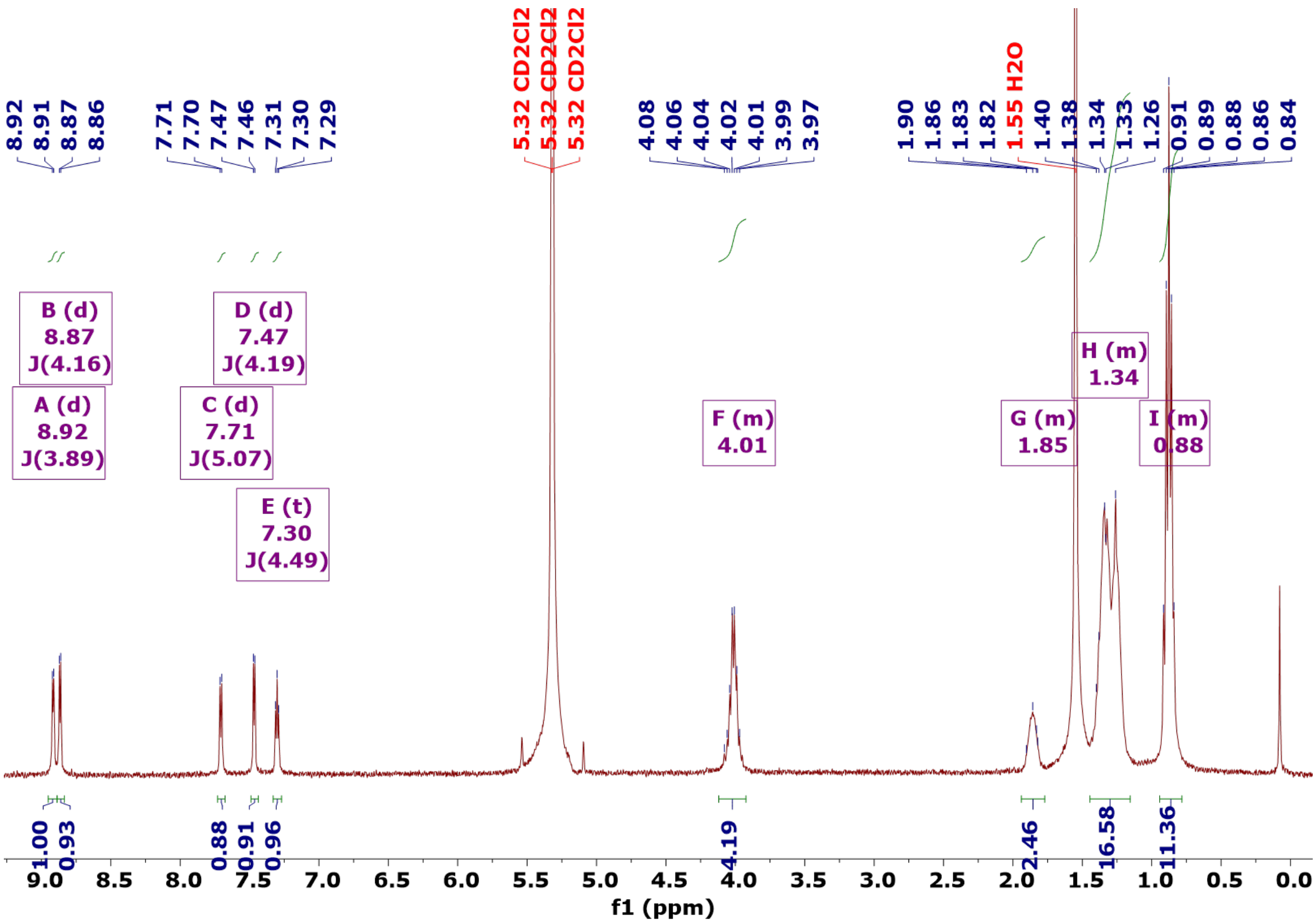

**Fig. S14** $^{1}$H NMR spectrum of TDPP-TDPP

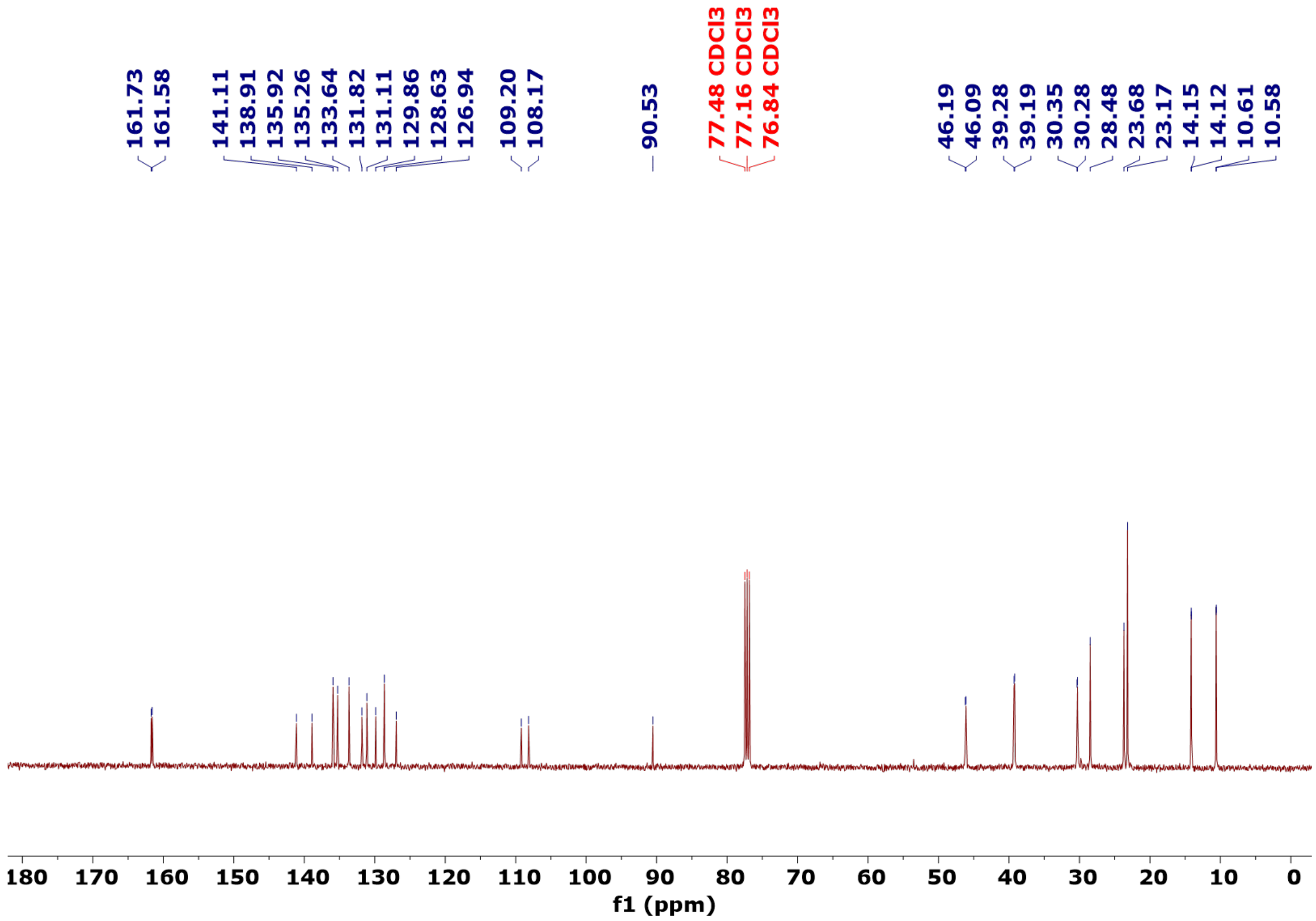

**Fig. S15** $^{13}$C NMR spectrum of TDPP-TDPP

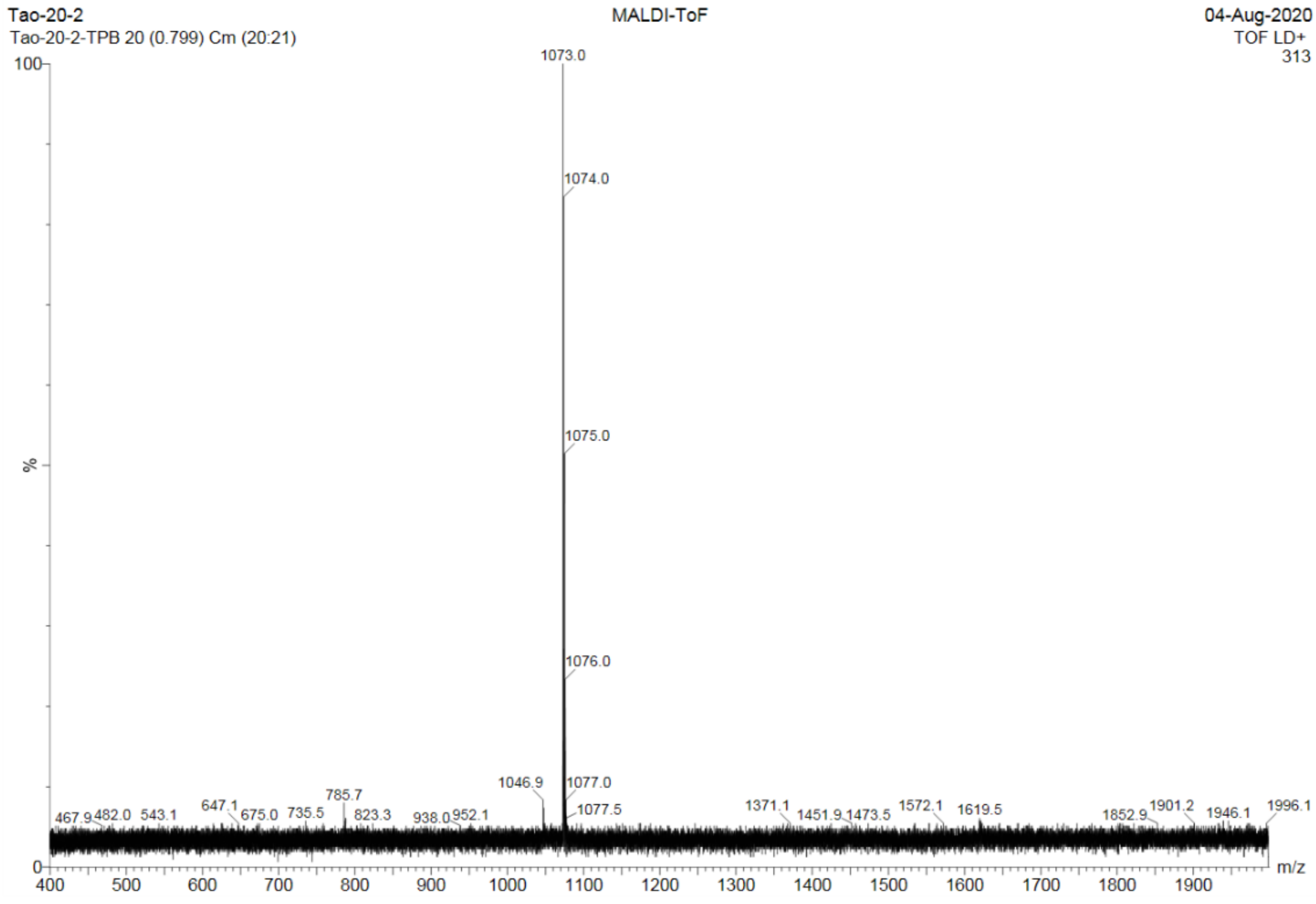

**Fig. S16** Mass spectrum of TDPP-v-TDPP

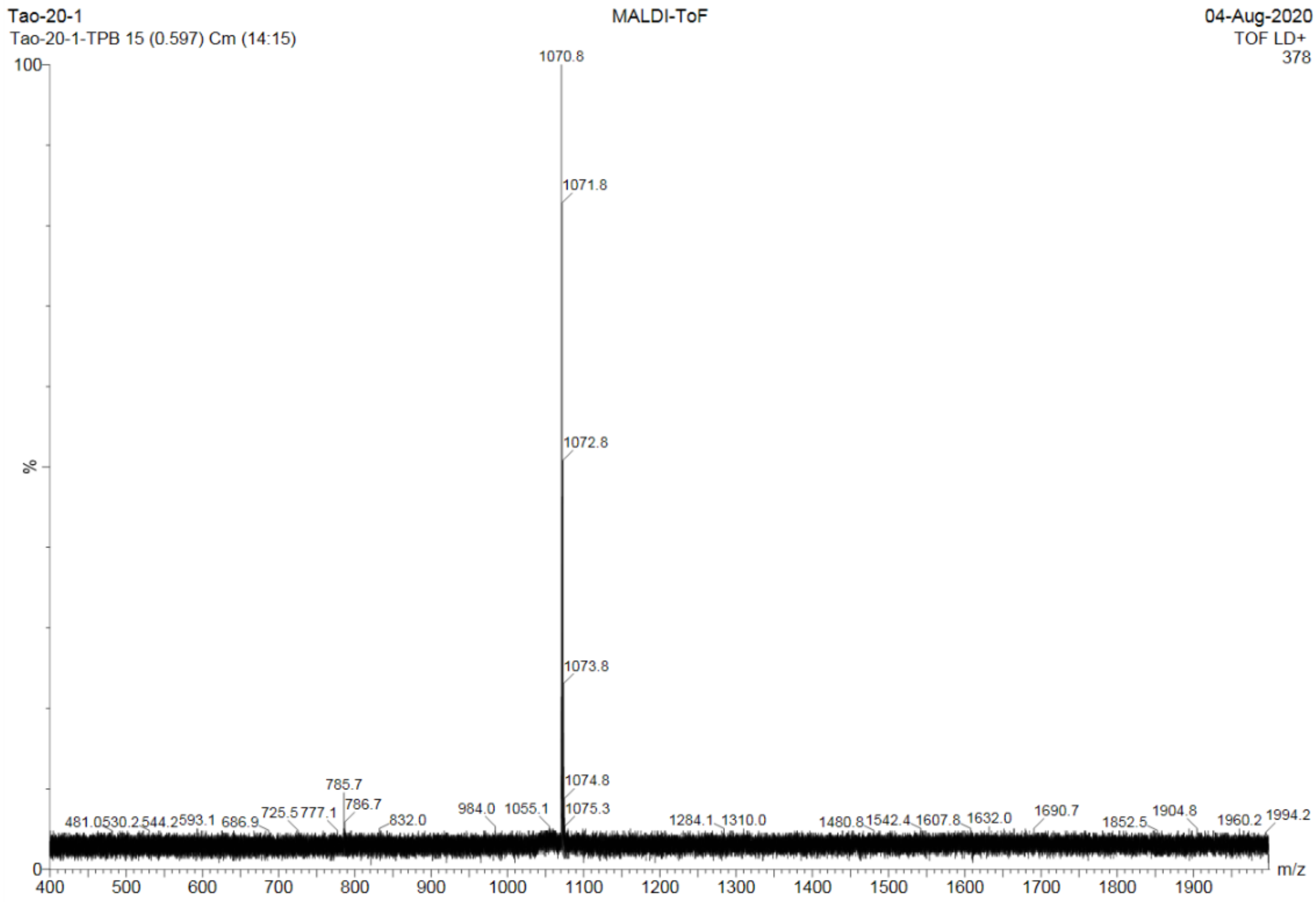

**Fig. S17** Mass spectrum of TDPP-TDPP

## 7. DFT Calculations

To reduce the computational complexity of both DFT/TD-DFT calculations, the effect of truncating the N-alkyl chains on the ground state geometry was investigated. These calculations used the B3LYP functional and a 6-311G(d,p) basis set.

The following substituents were trialled for TDDP-Br: *R* = –H (hydrogen), $–CH_3$ (methyl), –$CH_2CH(CH_3)_2$ (isobutyl), and $–CH_3$, $–CH_2CH(CH_3)_2$ for TDPP-v-TDPP. The calculations were not intended to be exhaustive but to inform whether the geometries associated with short-chain alkyl substituents were representative of the full molecular structure, and to help determine the symmetry point group of molecules. Conformations enforcing planar $C_{2h}$ geometries as well as those without any symmetry constraint were trialled. The latter were found to optimise into structures within the $C_i$ point group (inversion symmetry), where the terminal thiophene moieties rotated out of the molecular plane (as determined by the planar DPP core) by dihedral angle α- see Fig. S18.

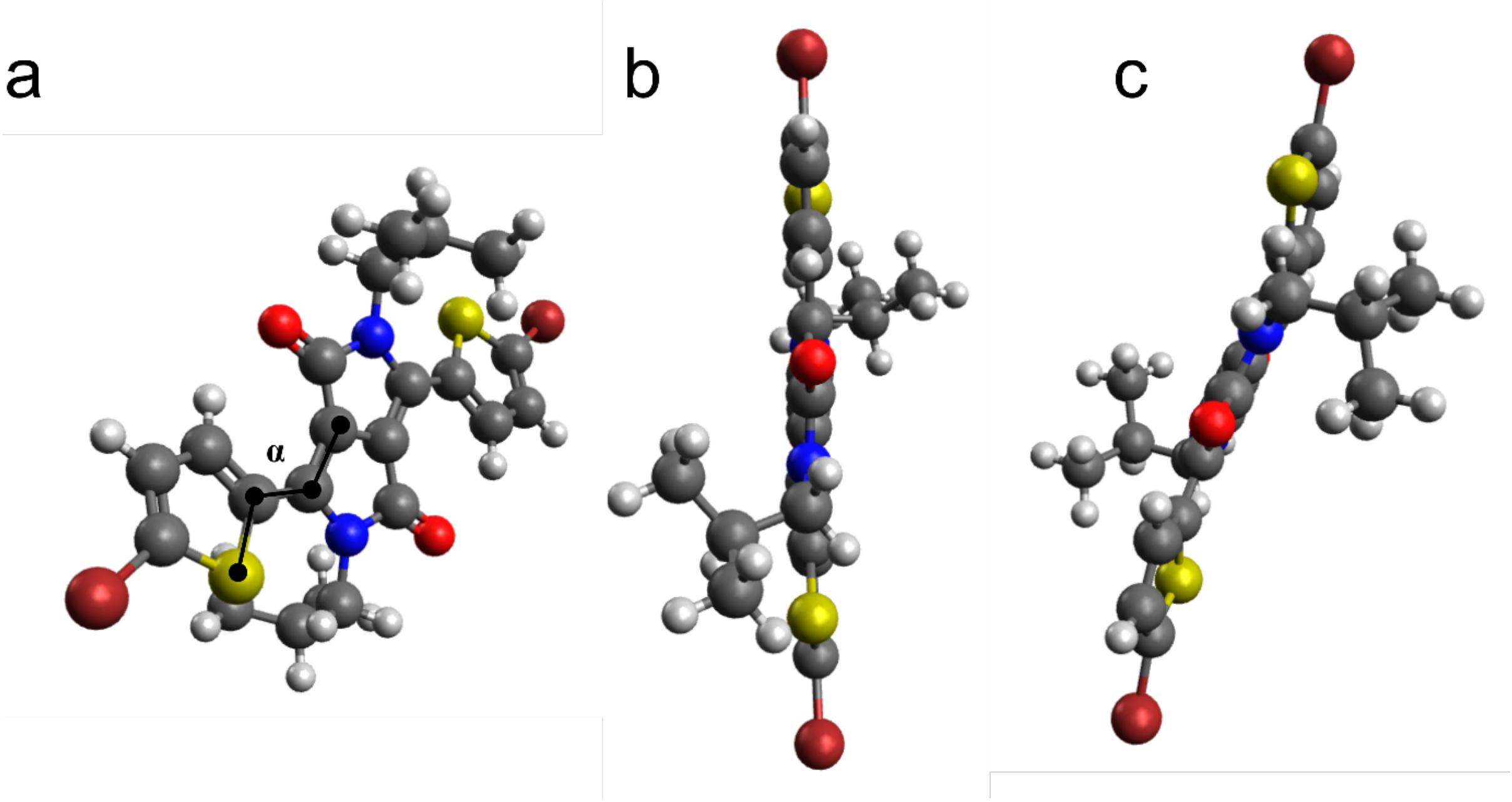

**Fig. S18** (a) TDDP-Br structure with illustration of dihedral angle (α), which is defined as the angle the thiophene rings make with the plane defined by the central DPP core. Example structures of α = (b) 0° and (c) 10°.

For the monomer and *R* = –H, calculations predicted the global ground state minimum energy geometry to be planar and $C_{2h}$ (see Table S3). Upon changing the *R* substituent to $–CH_2CH(CH_3)_2$, a sidechain group that is more representative of the true branched alkyl chains (see Fig. 1 in main paper), an added complication arises from the conformational complexity associated with the isobutyl substituents. In geometries where the hydrogens on the $–CH_2$ group of the isobutyl substituents point towards the thiophene ring (Fig. S19(a,b)), the terminal thiophene rings twist away from the planar DPP core in opposite directions ($\alpha$ = 32°), thus forming a structure which has overall $C_i$ symmetry. While, if the alkyl chains are rotated to avoid steric interactions with the aromatic core of the molecule (Fig. S19(c,d)), $\alpha$ reduces to 6 °. Locking this structure to planarity ($C_{2h}$) and re-optimisation returns a ground state minimum structure with an associated minimum energy only 11 meV above that of the twisted $C_i$ geometry ($\alpha$ = 6 °). Thermal energy at 298 K is 26 meV, and therefore it is likely that the ground state geometry will quickly interconvert (depending on the barrier height) between these two (and potentially other) minima. Thus, in a time-averaged picture, the planar $C_{2h}$ minima is representative of ground state TDPP-Br.

Calculations for the dimer (full results detailed in Table S3) with methyl and isobutyl *R* groups confirmed similar behaviour to the monomer: the $C_{2h}$ minimum (*R* = –isobutyl) was calculated to be 0.8 meV above that of the global $C_i$ minimum energy geometry- a value which is insignificant compared to solvent fluctuations or thermal energy at 298 K, again justifying approximating a $C_{2h}$ structure for TDPP-v-TDPP.

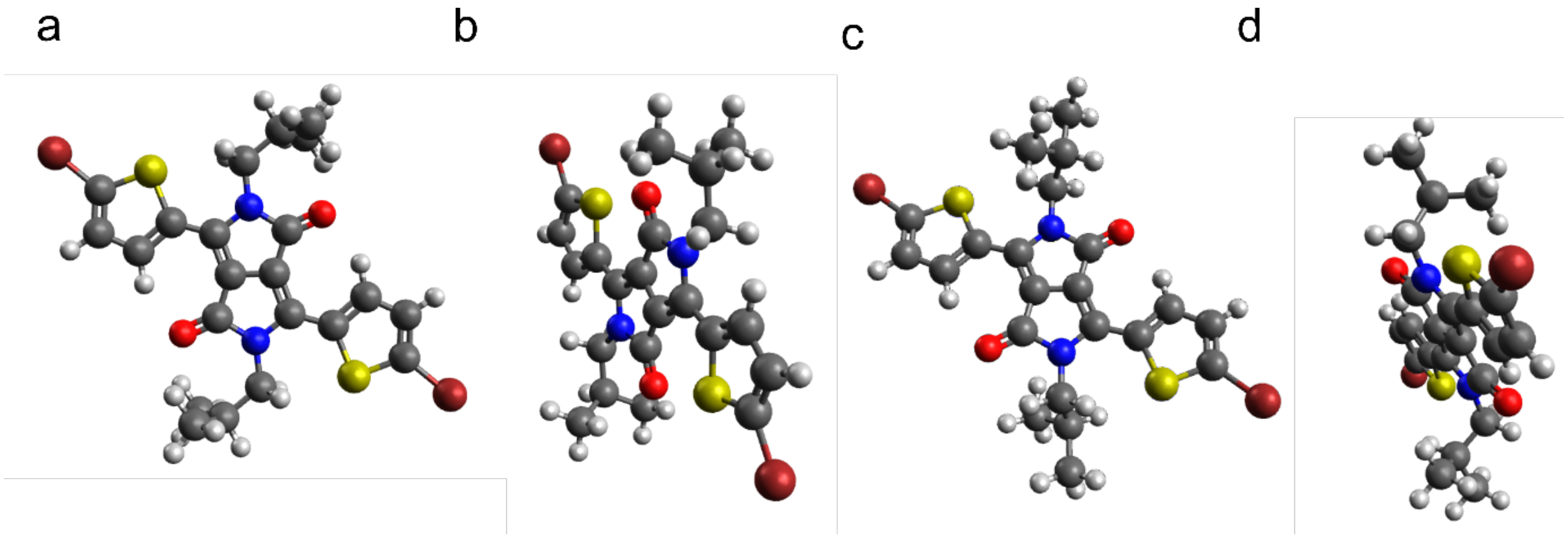

**Fig. S19** Two example TDPP-Br geometries (*R* = –isobutyl). (a,b) top-down and side-on perspectives of minimised geometry where the isobutyl side chain $CH_2$ hydrogens point to the terminal thiophene rings ($\alpha$ = 32°). (c,d) top-down and side-on depictions of the minimum energy where alkyl side chains are rotated to avoid steric interactions ($\alpha$ = 6°).

**Table S3.** Effect on conformation and side-chain length on TDPP-Br and TDPP-v-TDPP ground state energy and structure. Relative energies are grouped as a function of side chain functionalisation and molecule.

| Molecule | Side Chain Functionalisation | Ground state minimum energy / Hartree | Relative Energy / eV | Point Group | α / ° |
|---|---|---|---|---|---|
| TDPP-Br | Hydrogen | –6736.841270 | 0.000 | $C_{2h}$ | 0 |
| TDPP-Br | Methyl | –6815.464659 | 0.000 | $C_{2h}$ | 0 |
| TDPP-Br | Isobutyl[1] | –7056.508249 | 0.755 | $C_{2h}$ | 0 |
| TDPP-Br | Isobutyl[1] | –7056.521239 | 0.402 | $C_i$ | 32 |
| TDPP-Br | Isobutyl[2] | –7056.535996 | 0.000 | $C_i$ | 6 |
| TDPP-Br | Isobutyl[2] | –7056.535599 | 0.011 | ~$C_{2h}$ | 0 |
| TDPP-v-TDPP | Methyl | –3423.266994 | 0.000 | $C_{2h}$ | 0 |
| TDPP-v-TDPP | Isobutyl[1] | –3895.102568 | 0.0566 | $C_{2h}$ | 0 |
| TDPP-v-TDPP | Isobutyl[1] | –3895.142441 | 0.0167 | $C_i$ | 20 |
| TDPP-v-TDPP | Isobutyl[2] | –3895.159159 | 0.0000 | $C_i$ | 4 |
| TDPP-v-TDPP | Isobutyl[2] | –3895.158347 | 0.0008 | ~$C_{2h}$ | 0 |

[1,2] Alkyl chain conformations shown in Figure S21(a,b), and S21(c,d) respectively.

TD-DFT calculations (again using B3LYP/6-311G(d,p)) returned similar results for the excited bright state ($1B_u$) with isobutyl substituents, where the $C_{2h}$ structure was 14 meV higher in energy than the $C_i$ minima. Calculations for the dimer returned a 30 meV energy difference.

These global minimum energy geometries were used for vibrational normal mode calculations used to assign observed wavenumbers from the wavepacket FT analysis. For these vibrational calculations, $R$ = –$CH_3$ was used to minimise the computational cost. For TDPP-Br and TDPP-v-TDPP, the vibrational normal modes and associated wavenumbers the $S_0$ ($1A_g$) and bright excited ($1B_u$) electronic states were calculated using DFT and TD-DFT, respectively. The ground state minimum energy geometries were optimised and vibrational wavenumbers calculated with the DFT exchange-correlation functional B3LYP and a 6-311G(d,p) basis set.

The same basis function and basis set were used with TD-DFT to calculate the associated vibrational wavenumbers of the $1B_u$ excited states.

## 8. Assignment of vibrational wavepacket wavenumbers

**Table S4.** Assignment of TDPP-Br vibrational wavepacket modes present in the $S_0$ trace displayed in Fig. 4.

| Wavenumber / cm$^{-1}$ | Associated nuclear motions |
|---|---|
| 161 | Symmetric in-plane rocking of the terminal thiophene rings and the DPP core |
| 288 | Symmetric in-plane rocking of the terminal thiophene rings relative to the DPP core |
| 372 | Symmetric in-plane stretching of both C-Br bonds coupled to a horizontal sheering of the pyrrole rings through the central double bond |
| 505 | Symmetric in-plane rocking of the terminal thiophene rings relative to the DPP core |
| 701 | Symmetric out-of-plane twisting of the pi-conjugated system with majority amplitude on the DPP core |
| 1071 | In-plane asymmetric C-N-C double bond stretching motions |
| 1369 | Symmetric C-N-C stretching motions on the central DPP unit coupled to thiophene ring breathing motions |
| 1525 | In-plane symmetric C-C double bond over the conjugated backbone with large amplitude on the DPP units |

**Table S5.** Assignment of TDPP-Br vibrational wavepacket modes associated with the $1B_u$ state shown in Fig. 4.

| Wavenumber / $cm^{-1}$ | Associated nuclear motions |
|---|---|
| 288 | Symmetric in-plane rocking of the terminal thiophene rings and the DPP core |
| 494 | Anti-symmetric out-of-plane twisting of the π-conjugated system |
| 705 | Symmetric out-of-plane twisting of the π-conjugated system with majority amplitude on the DPP core |
| 1071 | In-plane asymmetric C-N-C double bond stretching motions |
| 1362 | Symmetric C-N-C stretching motions on the central DPP unit coupled to thiophene ring breathing motions |
| 1416 | Symmetric C-C double bond stretching on the central DPP unit coupled to thiophene ring breathing motions |
| 1527 | In-plane symmetric C-C double bond over the conjugated backbone with large amplitude on the DPP units |

**Table S6.** Assignment of TDPP-Br vibrational wavepacket modes associated with the $2A_g$ transient shown in Fig. 4.

| Wavenumber / $cm^{-1}$ | Associated nuclear motions |
|---|---|
| 965 | Anti-symmetric C-S-C stretching mode driven by a large amplitude C-Br stretching motion |
| 1325 | Anti-symmetric DPP and thiophene ring breathing modes |
| 1365 | Symmetric C-N-C stretching motions on the central DPP unit coupled to thiophene ring breathing motions |
| 1410 | Symmetric C-C double bond stretching on the central DPP unit coupled to thiophene ring breathing motions |
| 1479 | Anti-symmetric stretching of the central C=C bonds on the DPP unit coupled to C=C stretching modes over the whole conjugated system. |

**Table S7.** Assignment of TDPP-v-TDPP $S_0$ vibrational wavepacket modes- see Fig. 6.

| **Wavenumber / cm$^{-1}$** | **Associated nuclear motions** |
|---|---|
| 215 | Symmetric in-plane rocking of the thiophene and the DPP rings |
| 614 | Symmetric thiophene ring breathing modes of the terminal thiophene |
| 835 | Symmetric thiophene and DPP ring breathing modes across the whole structure |
| 937 | Out-of-plane rocking of the hydrogens on the central thiophene rings and vinyl linker |
| 1162 | Anti-symmetric stretching of the C=C bonds on the central thiophene rings |
| 1416 | Symmetric thiophene and DPP ring breathing modes across the whole structure |

**Table S8.** Assignment of TDPP-v-TDPP $1B_u$ vibrational wavepackets, data shown in Fig. 6.

| Wavenumber / $cm^{-1}$ | Associated nuclear motions |
|---|---|
| 215 | Anti-symmetric in-plane rocking of the thiophene and the DPP rings |
| 1467 | In-plane C=C double bond stretching modes on the DPP units and terminal thiophene rings. |
| 1502 | In-plane C=C double bond stretching modes on the DPP units and terminal thiophene rings. |
| 1531 | In-plane C=C double bond stretching modes over the whole pi-conjugated system with larger amplitude on the terminal DPP/thiophene units |

**Table S9.** Assignment of TDPP-v-TDPP 2$A_g$ state vibrational wavepackets- see Fig. 6.

| Wavenumber / $cm^{-1}$ | Associated nuclear motions |
|---|---|
| 54 | In-plane asymmetric rocking of the DPP units relative to each other through the central thiophenes/vinyl linker. |
| 1107 | Low amplitude thiophene and DPP ring breathing motions on all rings in the molecular structure. |
| 1258 | High amplitude symmetric C-C double bond stretching modes on the vinyl linker and central thiophene rings. |
| 1422 | Symmetric and in-phase thiophene and DPP ring breathing motions on all rings in the molecular structure. |
| 1502 | In-plane C=C double bond stretching modes on the DPP units and terminal thiophene rings. |

**Table S10.** Assignment of DPPDTT $S_0$ vibrational wavepacket modes- see Fig. S5.

| Wavenumber / cm$^{-1}$ | Associated nuclear motions |
|---|---|
| 725 | In-plane asymmetric C-S-C stretching motion |
| 856 | Out-of-plane sheering of the DPP units via a symmetric C=C-C bending motion. |
| 946 | In-plane asymmetric stretching motion of the S-C=C-S bonds on the centre unit of the thiophene chain which drives a low amplitude ring breathing motion along the whole thiophene chain. |
| 1072 | Thiophene ring breathing mode on the terminal thiophene of the dimer unit |

**Table S11.** Assignment of DPPDTT $S_{bright}$ vibrational wavepacket modes- see Fig. S5.

| Wavenumber / cm$^{-1}$ | Associated nuclear motions |
|---|---|
| 727 | In-plane asymmetric C-S-C stretching motion coupled to in-plane rocking of the DPP units |
| 1304 | DPP and thiophene ring breathing motions throughout the whole molecular structure |
| 1445 | DPP and thiophene ring breathing motions throughout the whole molecular structure |

**Table S12.** Assignment of DPPDTT $S_{dark}$ vibrational wavepacket modes- see Fig. S5.

| Wavenumber / $cm^{-1}$ | Associated nuclear motions |
|---|---|
| 618 | Out-of-plane C=C-C twisting motions on the thiophene chain |
| 703 | In-plane asymmetric C-S-C stretching motion coupled to in-plane rocking of the DPP units |
| 734 | In-plane asymmetric C-S-C stretching motion coupled to in-plane rocking of the DPP units |
| 859 | Low amplitude thiophene and DPP ring breathing modes |
| 946 | In-plane asymmetric stretching motion of the S-C=C-S bonds on the centre unit of the thiophene chain which drives a low amplitude ring breathing motion along the whole thiophene chain. |